\documentclass[10pt,journal,final,twoside]{IEEEtran}

\IEEEoverridecommandlockouts
\usepackage{mysymbol}

\usepackage{xspace,empheq,fancybox,amsmath,amssymb,graphicx,epstopdf,epsfig,syntonly,times,amsthm}
\usepackage{psfrag,color,bm,array,cite}
\usepackage{url,cite,footnote,xspace,syntonly,algorithm,algorithmic}
\usepackage{verbatim,multirow}
\usepackage[T1]{fontenc}
\usepackage{mathtools}
\usepackage{amsmath,amsfonts,amsthm,bm} % Math packages
\usepackage{graphicx}
\usepackage{mwe}
\usepackage{caption}
\usepackage{cite}
\usepackage{amsmath,amssymb,amsfonts}
\usepackage{algorithmic}
\usepackage{graphicx}
\usepackage{textcomp}
\usepackage[caption=false,font=footnotesize]{subfig}
\usepackage{color}
\usepackage{xcolor}
\usepackage{cite}
\usepackage{graphicx}

\usepackage{epstopdf}
\usepackage{comment}
\usepackage{multirow}
\usepackage{paralist}
\usepackage{lineno}
\usepackage{mdwlist}
\usepackage{eurosym}\DeclareGraphicsExtensions{.pdf,.png,.jpg}
\usepackage{breqn}
\usepackage{makecell}
\usepackage{soul}
\usepackage{empheq}
\graphicspath{{./pic/}}
\usepackage[super]{nth}
\usepackage{subfig}
\usepackage{bm}
\usepackage[normalem]{ulem}
\usepackage{xcolor}
\def\BibTeX{{\rm B\kern-.05em{\sc i\kern-.025em b}\kern-.08em
    T\kern-.1667em\lower.7ex\hbox{E}\kern-.125emX}}

\newcommand\upj{\mathord{\mathrm{j}}}
\usepackage{accents}

\newtheorem{remark}{\bfseries Remark}

\newcommand{\yc}{\color{teal}{}}

\allowdisplaybreaks

\usepackage{pifont}% http://ctan.org/pkg/pifont

\theoremstyle{remark}

\DeclarePairedDelimiterX\set[1]\lbrace\rbrace{\def\given{\;\delimsize\vert\;}#1}

\begin{document}
\title{Data-driven Modeling of Linearizable Power Flow for Large-scale Grid Topology Optimization
%Generative Modeling of AC Power Flow: How to Efficiently Solve Grid Resilience Optimization
}
%\title{Resilient Grid Operations Based on the Piecewise Linearization of the AC Power Flow through Generative Modeling
%Generative Modeling of AC Power Flow: How to Efficiently Solve Grid Resilience Optimization
%}
\author{\IEEEauthorblockN{Young-ho Cho},~\IEEEmembership{Student Member, IEEE}, and \IEEEauthorblockN{Hao Zhu},~\IEEEmembership{Senior Member, IEEE}

\thanks{\protect\rule{0pt}{3mm} This work has been supported by NSF Grants 2130706 and 2150571.} \thanks{\protect\rule{0pt}{3mm} The authors are with the Chandra Family Department of Electrical \& Computer Engineering, The University of Texas at Austin, Austin, TX, 78712, USA; Emails: {\{jacobcho, haozhu\}{@}utexas.edu}.}
}
\maketitle

\begin{abstract}
    Effective power flow (PF) modeling critically affects the solution accuracy and computational complexity of large-scale grid optimization problems. Especially for grid optimization involving flexible topology to enhance resilience, obtaining a tractable yet accurate approximation of nonlinear AC-PF is essential.
    This work puts forth a data-driven approach to obtain piecewise linear (PWL) PF approximation using an innovative neural network (NN) architecture, effectively aligning with the inherent generative structure of AC-PF equations.
    Accordingly, our proposed generative NN (GenNN) method directly incorporates binary topology variables, efficiently enabling a mixed-integer linear program (MILP) formulation for grid optimization tasks like optimal transmission switching (OTS) and restoration ordering problems (ROP).
    To attain model scalability for large-scale applications, we develop an area-partitioning-based sparsification approach by using fixed-size areas to attain a linear growth rate of model parameters, as opposed to the quadratic one of existing work. 
    Numerical tests on the IEEE 118-bus and 6716-bus synthetic Texas grid demonstrate that our sparse GenNN achieves superior accuracy and computational efficiency, substantially outperforming existing approaches in large-scale PF modeling and topology optimization.
\end{abstract}

\section{Introduction}

%Grid resilience enhancements
%ACOPF
%Nonconvex quad. equality AC PF
%Most problems are the extension of AC OPF.
%How extended?

%%%%%%%%%%%%%%%%%5
\IEEEPARstart{P}{ower} flow (PF) modeling is an essential power system analysis task for attaining efficient and reliable grid operations.
PF-based grid optimization is indispensable for enhancing the flexibility and resilience of power grids under uncertain and extreme operating conditions.
The nonlinearity of AC-PF equations greatly challenges grid optimization tasks including the clairvoyant optimal power flow (OPF). Recent interest in grid resilience has led to optimal transmission switching (OTS) \cite{fisher2008optimal} and restoration ordering problem (ROP) \cite{van2011vehicle} of increased complexity due to the integer topology variables therein.
Hence, an accurate yet tractable approximation of the nonlinear AC-PF model is imperative for improving the optimality and computation performance for solving grid optimization problems, especially those with topology variables.

%Linearized PF and model-based PF (can be the next paragraph)
%DC, lin AC, piecewise lin AC

%Developing PF approximation models has been an active research area. 
%Notably, linearized power flow models have been advocated for their simplicity, such as the well-known DC model~\cite{stott2009dc} and the more general first-order approximation at a given operating point for improved accuracy~\cite{coffrin2014linear}.
%However, linear models are very limited in their generalizability across all possible operating conditions~\cite{trodden2013optimization}.
%A straightforward extension is the piecewise linear (PWL) approximation, which utilizes multiple operating points for linearization; see e.g.,~\cite{brown2020transmission}. 
%There is a trade-off between accuracy and complexity in these model-based approaches, and the best number and location of operating points can be difficult to determine.
%To tackle these issues, data-driven approaches have been recently developed as an alternative to PF approximation.

The data-driven PF approximation has recently become popular thanks to advances in machine learning (ML) techniques.
Similar to data-driven PF linearization \cite{liu2018data,buason2025sample} developed to increase accuracy over classical linearized PF models at \textit{fixed} points, ML models trained from realistic PF scenarios can improve over traditional piecewise-linear (PWL) approximation based on a rigid list of linearized regions. Recent work of PWL-PF models is based on $K$-plane regression~\cite{chen2021data} and neural networks (NNs) \cite{kody2022modeling,chevalier2022global}, where
the choice of ReLU activation  can construct simple yet accurate PWL models by incorporating the PF Jacobian information. Similar ideas have been explored in the constraint learning framework \cite{chen2023efficient,sang2023encoding} even for non-PF constraints. ReLU-based PWL modeling is particularly attractive for grid optimization by allowing the reformulation into mixed-integer linear programs (MILPs), for which efficient off-the-shelf solvers exist.

\begin{comment}
Despite recent advances in data-driven PF models, existing approaches still fail to address scalability needs when applied to large systems, particularly for grid topology optimization tasks. As detailed in Table \ref{tbl:literature}, methods such as those in \cite{kody2022modeling}, \cite{chevalier2022global}, and \cite{jalving2024physics}  exhibit a quadratic growth $\ccalO(N^2)$ in NN parameters, leading to increased computational burdens, extensive data requirements, and prolonged training times as system size expands. Although alternative strategies like those in \cite{liu2018data}, \cite{buason2025sample}, and \cite{lin2024powerflownet} achieve linear $\ccalO(N)$ scaling, they lack key attributes necessary for grid optimization. This discrepancy highlights the need for \textit{low-complexity} NN-based PWL-PF models that maintain high approximation accuracy while ensuring a tractable MILP reformulation, thereby making them more applicable for large-scale systems. 
\end{comment}

Despite recent advances in data-driven PF models, existing approaches still fail to address scalability needs for large practical systems, particularly for grid topology optimization tasks. See a detailed comparison in Table \ref{tbl:literature}.
As the system size increases, the model parameter complexity grows rapidly, imposing significant computational burdens and data overfitting issues.
While simple linearized models \cite{liu2018data,buason2025sample} tend to have the parameters in $(\ccalO(N))$, their approximation accuracy is low. To improve modeling accuracy, almost all existing NN-based methods \cite{kody2022modeling,chevalier2022global,jalving2024physics,cho2023topology} under fully-connected layers inevitably suffer from a quadratic growth $(\ccalO(N^2))$, thereby limiting their applications to large systems. A recent work \cite{lin2024powerflownet} has used graph-embedded layers to achieve linear scaling. However, it is strongly questionable whether the complex layer designs in \cite{lin2024powerflownet} are compatible with grid optimization tasks. 
Hence, \textit{low-complexity} NN-based PF models are still missing for large-scale applications to grid optimization.

Moreover, existing ML models primarily focus on PF prediction under a fixed topology and are unable to conveniently adapt to new topology. This limitation is especially critical for solving optimization tasks such as OTS and ROP for increased grid resilience and flexibility, while all possible topology scenarios have to be accommodated.
This need is clearly demonstrated by recent work on topology-adaptive learning-for-OPF, using topology-informed input features \cite{zhou2022deepopf} or a graph-topology embedding design \cite{liu2022topology}.  However, existing data-driven PF models have rarely considered \textit{a variable topology}. The only exception is the PF model using topology embeddings in \cite{lin2024powerflownet}, and yet it  requires to generate training data samples under the exhaustive list of all possible topology scenarios. This approach not only significantly increases the data generation time, but is also unable to incorporate an unseen topology, making it not scalable for grid topology optimization. In summary, it is imperative to develop NN-based PF approximation models for solving large-scale grid topology optimization tasks in an efficient manner.

To this end, we seek to develop a new data-driven approach to linearizable PF modeling that can attain high PF consistency and low parameter complexity, which are especially important to the ensuing large-scale topology optimization tasks. The key of our design is the \textit{generative NN structure}, or the GenNN model, that follows the PF physics of nodal power balance and embeds the binary line-status variables in the layers. The resultant PWL-PF approximation naturally maintains consistency across the predicted PF variables and integrates flexible topology decisions. Another key feature of our work is the \textit{low-complexity GenNN design} for scalability,  by adopting an area-based sparsification strategy to partition large systems into smaller, fixed-size areas via spectral clustering. As a result, the number of ReLU activation neurons scales linearly with the system size, as opposed to the typical quadratic growth in existing approaches. Note that this scalable model design can significantly reduce the number of ReLU-based binary variables, and thus greatly enhance the efficiency of solving the MILP reformulation of the ensuing grid optimization tasks. Our GenNN has successfully applied to topology optimization tasks like  OTS and ROP, demonstrating its excellent accuracy and computational efficiency.
Its improvements in PF consistency and accuracy over existing direct NN approaches in \cite{kody2022modeling,chevalier2022global,jalving2024physics} have been validated using the IEEE 118-bus test case~\cite{IEEE_case_ref} and the 6716-bus synthetic Texas grid case~\cite{mateo2024building}. Notably, we are the first to demonstrate the applicability of NN-based PWL models for solving AC-OTS/ROP typed grid optimization in large-scale systems like the 6716-bus Texas grid.
In short, our main contributions are three-fold:
\begin{enumerate}
    \item We develop a data-driven GenNN model for accurate PWL-PF approximation, that can explicitly ensure the consistency among PF variables and topology adaptivity.
    \item We put forth a new low-complexity model design using  area-based partitioning, that can attain a \textit{linear} parameter complexity order  for large-scale system applications.
    \item Our GenNN attains topology adaptivity for large-scale grid topology optimization by embedding binary line status into the layers, easily modeling any unseen topology without exhaustive generation of data samples.
\end{enumerate}

The rest of the paper is organized as follows. Section~\ref{sec:lin} presents the AC-PF modeling. Section~\ref{sec:piecewise} discusses the GenNN model and its low-complexity design using sparsification.
In Section~\ref{sec:MILP_opt}, we develop the steps for MILP formulation by simplifying the equivalent constraints.
Section~\ref{sec:sim} presents the results to demonstrate the performance improvements attained by our models and reformulation over the existing direct approaches, and Section~\ref{sec:con} wraps up the paper.

\begin{comment}
\begin{enumerate}
    \item We develop a data-driven GenNN model that can provide accurate PWL approximation of AC-PF while maintaining the consistency among different  PF variables; 
    \item We design the sparse GenNN to reduce the model complexity by exploiting the weak PF coupling among areas, that can easily scale up to large interconnections.
    \item Our GenNN model includes a varying grid topology as decision variables, enabling an MILP-based reformulation of a variety of grid topology optimization tasks.
\end{enumerate}
\end{comment}

\begin{table}[t!]
\begin{center}
\caption{Comparisons of data-driven PF models.}
\label{tbl:literature}
\begin{tabular}{c|c|c|c|c}
\hline
Study & \makecell{Parameter\\complexity} & \makecell{Topology\\adaptivity} & \makecell{Optimization\\compatibility} & \makecell{System\\size}\\
\hline
\cite{liu2018data}          & $\ccalO(N)$   & $\times$       & $\checkmark$    & 123 buses \\
\cite{buason2025sample}      & $\ccalO(N)$   & $\times$       & $\checkmark$   & 2,383 buses \\
\cite{kody2022modeling}      & $\ccalO(N^2)$ & $\times$       & $\checkmark$   & 89 buses \\
\cite{chevalier2022global}   & $\ccalO(N^2)$ & $\times$       & $\checkmark$   & 200 buses \\
\cite{jalving2024physics}    & $\ccalO(N^2)$ & $\times$       & $\checkmark$   & 118 buses \\
\cite{cho2023topology}       & $\ccalO(N^2)$ & $\checkmark$   & $\checkmark$   & 118 buses \\
\cite{lin2024powerflownet}   & $\ccalO(N)$   & $\times$       & ? %$\times$       
& 6,470 buses \\
Our work                   & $\ccalO(N)$   & $\checkmark$   & $\checkmark$   & 6,716 buses \\
\hline
\end{tabular}
\end{center}
\end{table}

\section{AC Power Flow Modeling} \label{sec:lin}
We first present the nonlinear AC-power flow (PF) modeling to be used by our proposed data-driven model later on.
Consider a transmission system consisting of $N$ buses collected in the set $\mathcal{N} := \{1,\dots, N\}$ and $L$ branches in $\mathcal{L} := \{(i, j)\} \subset \mathcal{N} \times \mathcal{N}$.
For each bus $i\in\mathcal{N}$, let $V_i\angle\theta_i$ denote the complex nodal voltage phasor, and  $\{P_i,Q_i\}$ denote the active and reactive power injections, respectively.
For each branch $(i,j)\in\mathcal{L}$, let $\theta_{ij}:= \theta_i-\theta_j$ denote the angle difference between buses $i$ and $j$, with $\{P_{ij},Q_{ij}\}$  the active and reactive power flows from bus $i$ to $j$.
In addition, the branch's series and shunt admittance values are respectively denoted by $y_{ij}= g_{ij}+\upj b_{ij}$ and $y_{ij}^{\mathrm {sh}}=g^{\mathrm {sh}}_i+\upj b^{\mathrm {sh}}_i$.

The power flows of  branch $(i,j)$ depend on $\{V_i, V_j,\theta_{ij}\}$, and in the case of a transformer, its tap ratio $a_{ij}$, as given by
\begin{subequations}
\label{powerflow}
\begin{align}
    P_{ij}\!&=\!V^2_{i} \big(\frac{ g_{ij}}{ a^2_{ij}}\!+\!g^{\mathrm {sh}}_i\big)\!-\! \frac{ V_{i}  V_{j}}{ a_{ij}} \big( g_{ij}\cos \theta_{ij}\!+\!b_{ij}\sin \theta_{ij}\big), \label{powerflow_1}\\
    Q_{ij}\!&=\! - V^2_{i} \big(\frac{ b_{ij}}{a^2_{ij}}\!+\!b^{\mathrm {sh}}_i\big) \! -\! \frac{ V_{i}  V_{j}}{ a_{ij}} \big( g_{ij}\sin  \theta_{ij} \!- \!b_{ij}\cos  \theta_{ij}\big).
\end{align}
\end{subequations}
The following nonlinear terms can capture the coupling between active and reactive power flows in \eqref{powerflow}: 
\begin{align}
    \gamma_i := V^2_i,~\rho_{ij}:= V_i  V_j \cos  \theta_{ij},~\mathrm{and}~\pi_{ij}:= V_i  V_j \sin \theta_{ij}. \label{eq:nl}
\end{align}
Note that both power flows are linear combinations of these nonlinear terms, based on the branch parameters $\{y_{ij}, y^{\mathrm {sh}}_{ij}, a_{ij}\}$.
For transmission lines,  $a_{ij}$ is simply set to 1. For transformers, the tap ratio $a_{ij}$ is typically within the range of $[0.9, 1.1]$ and only affects the primary-to-secondary direction.  Thus, for the secondary-to-primary direction, one can also use $a_{ij}=1$ in \eqref{powerflow}. While this work assumes $a_{ij}$ is given, it is possible to include controllable tap ratios as optimization's decision variables; see e.g.,~\cite{bazrafshan2018optimal}.

The power injections to bus $i$ can be formed by $\{P_{ij}, Q_{ij}\}$ and the corresponding branch status denoted by $\epsilon_{ij} \in \{0,1\}$ (0/1: off/on), as given by 
%Then, the nodal power balance at bus $i$ becomes
\begin{subequations}\label{powerinj}
\begin{align}
P_{i} &= \textstyle \sum_{(i,j) \in \mathcal{L} } ~\epsilon_{ij} P_{ij},\\
Q_{i} &= \textstyle \sum_{(i,j) \in \mathcal{L} } ~\epsilon_{ij} Q_{ij}.
\end{align}
\end{subequations}
The binary variables $\{\epsilon_{ij}\}$ will be important for formulating grid optimization tasks with varying topology, as detailed later on. With no topology changes, they can be fixed at $\epsilon_{ij}=1$.

One can view the above AC-PF model as a sequential transformation from the angles and voltages to i) the nonlinear terms, ii) line power flows, and iii) bus injections.
To represent this transformation, let us concatenate all the power flows and injections in $\bbz^{pf} \in \mathbb{R}^{4L}$ and $\bbz^{inj}\in \mathbb{R}^{2N}$.
In addition, let $\bbgamma \in \mathbb{R}^{N}$, $\bbrho \in \mathbb{R}^{L}$, and $\bbpi \in \mathbb{R}^{L}$ denote the vector-forms of the three types of nonlinear terms.
This way, we have 
\begin{subequations}\label{pow}
\begin{align}
    \bbz^{pf} &= \bbW^{\gamma} \bbgamma + \bbW^{\rho} \bbrho + \bbW^{\pi} \bbpi,\label{powa}\\
    \bbz^{inj} &= \bbW^{\psi} \bbz^{pf}\label{powb}
\end{align}
\end{subequations}
where the weight matrices $\{\bbW^{\gamma}, \bbW^{\rho}, \bbW^{\pi}, \bbW^{\psi}\}$ are of appropriate dimensions using the given branch parameters and branch status variables $\epsilon_{ij}$ in \eqref{powerflow} and \eqref{powerinj}.
%Nonlinearity: more detail, only nonlinearity comes from the nonlinear terms
Clearly, the nonlinear terms make the rest of transformations in \eqref{powa}-\eqref{powb} purely linear. 
Thus, our proposed NN model will represent \eqref{pow} as linear layers and use the nonlinear layers only to form $\{\bbgamma,\bbrho,\bbpi\}$.% in \eqref{eq:nl}. 

We also discuss the standard PF linearization around a given operating point, which will be used for the design of PWL approximation.
We consider the first-order approximation method, while there also exist other linearization techniques such as the fixed-point method~\cite{simpson2017theory}. Consider the vector $\bbx$ consisting of all input variables in $\{V_i\}_{i\in\mathcal N}$ and  $\{\theta_{ij}\}_{(i,j)\in\mathcal{L}}$, and a given operating point denoted by $\check{\bbx}$ along with  $\{\check V_i,\check \theta_{ij}\}$.
To linearize $\bbgamma$, we can use $\hat \gamma_i=  2V_i - \check V_i$, $\forall i\in\mathcal{N}$, by using the fixed $\check V$. As shown in~\cite{trodden2013optimization}, for power systems with well-regulated bus voltages this linearized approximation %of $\gamma_i, \forall i\in\mathcal{N}$, denoting as 
vector $\hat {\bbgamma}$ can attain very high accuracy.
As for the two nonlinear terms in $[\bbrho;~\bbpi] = \bbf(\bbx) \in \mathbb R^{2L}$, where $\bbf(\cdot)$  represents the nonlinear mapping described in \eqref{eq:nl}.
For given $\check {\bbx}$, the first-order approximation becomes~\cite{cho2023topology}
\begin{align}
 \hhatbbgamma = 2\bbV - \check{\bbV},~~\textrm{and}~~[\hhatbbrho; \hhatbbpi]%&= \bbf(\check {\bbx})+ \check{\bbJ} (\bbx - \check {\bbx}) \nonumber\\
    &= \bbf(\check {\bbx})+ \check{\bbJ} \tilde{\bbx} \label{linearf}
\end{align}
where $\check{\bbJ}$ denotes the Jacobian matrix of $\bbf(\cdot)$ evaluated at $\check {\bbx}$, while  $\tilde{\bbx} := \bbx -\check {\bbx}$ stands for the deviation from $\check {\bbx}$. Due to the sinusoidal relations, the linearization in $[\hhatbbrho; \hhatbbpi]$ could suffer from high inaccuracy issue, especially with a large angle difference. 
Therefore, the ensuing section will build the PWL model to improve the approximation of $[\bbrho; \bbpi]$.
%To intuitively improve the approximation accuracy, we can build the piecewise linear (PWL) model with multiple operating points.
%The PWL model for $\bbrho$ and $\bbpi$ replaces the nonlinear mapping in the first step, so the resultant model still retains the structure in \eqref{pow}.

\begin{figure}[t!]
	\begin{center}
		\includegraphics[scale=0.33]{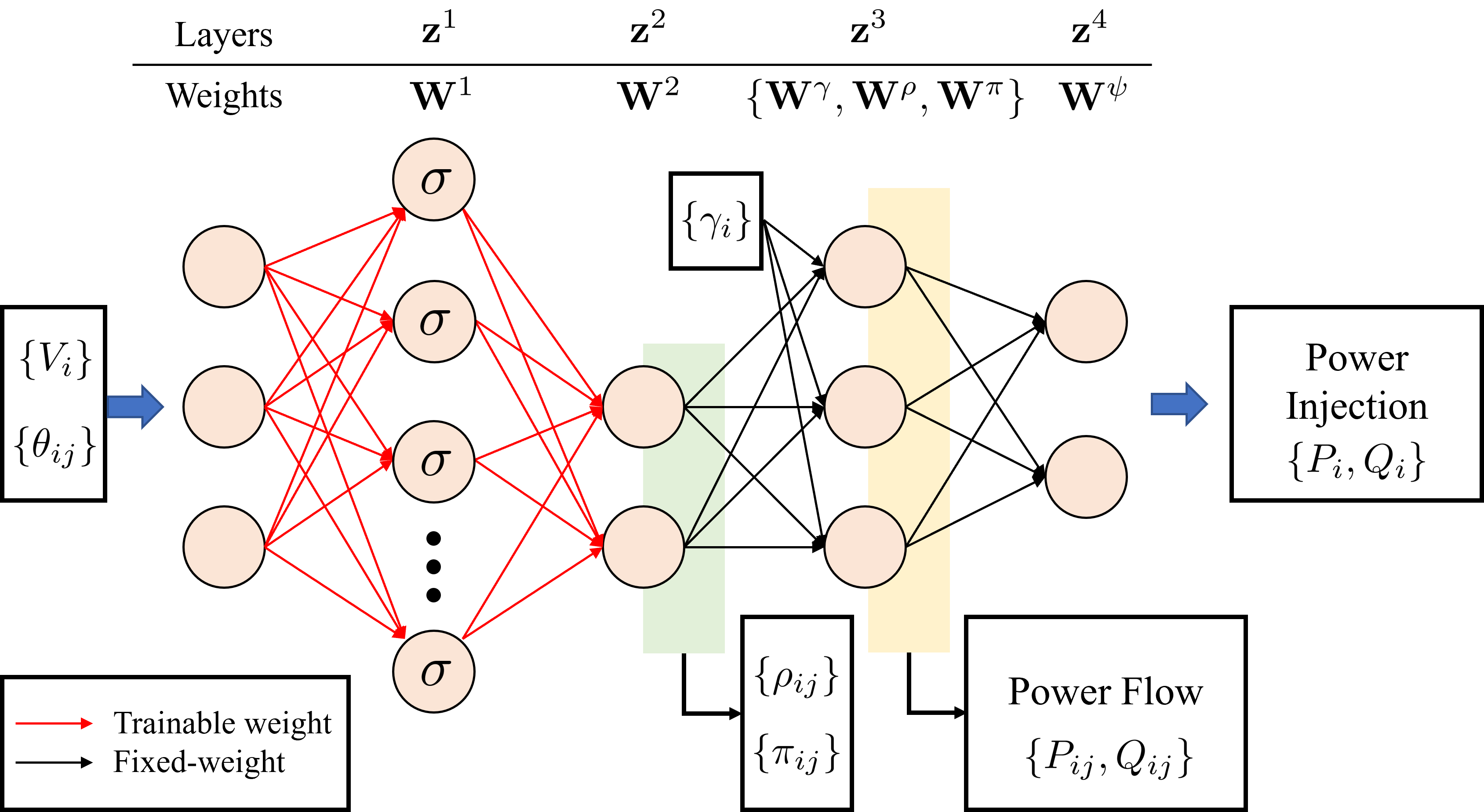}
		\caption{\small The proposed GenNN model predicts the nonlinear terms in the second layer using trainable weights; it further generates power variables in the third/fourth layers to match the PF equations.} \label{structure}
	\end{center}
\end{figure}

\section{GenNN for PWL-PF Approximation}
\label{sec:piecewise}
We develop the PWL model of PF equations by using a generative neural network (GenNN) structure.  Fig.~\ref{structure} illustrates the overall architecture of our proposed design. The first two layers predict the nonlinear terms $[\bbrho;\bbpi]$, and the following two \textit{linear} layers generate the power flows and injections.
The ReLU activation $\sigma(\cdot)$ in the first (hidden) layer can lead to a PWL model for $[\hhatbbrho; \hhatbbpi]$ while adjusting the GenNN parameters therein can improve the accuracy of the approximation.
The full structure of the GenNN becomes:
\begin{subequations}\label{NN_pf}
\begin{align}
    \bbz^{1} &= \sigma(\bbW^{1} \tdbbx + \bbb),\label{NN_pf1}\\
    \bbz^{2} &= \bbf(\check{\bbx}) + \Big( \check{\bbJ} \tdbbx +  \bbW^2 \bbz^{1} \Big),\label{NN_pf2}\\
    \bbz^{3} &= \bbW^{\gamma} \hhatbbgamma + [\bbW^{\rho};\bbW^{\pi}] \bbz^{2},\label{NN_pf3}\\
    \bbz^{4} &= \bbW^{\psi} \bbz^{3}.\label{NN_pf4}
\end{align}
\end{subequations}
\begin{remark}\label{rmk:layer} (GenNN layer design.)
\rm Our GenNN is designed such that each layer produces the subset of PF variables per the underlying sequential relations of the full PF model. 
The first two layers use an NN-based PWL approximation to generate the key nonlinear terms, where the use of Jacobian matrix of an operating point $\check{\bbx}$ follows from  \cite{kody2022modeling,chevalier2022global, cho2023topology}. Accordingly, the third layer produces line power flows, while the fourth layer the bus injections, both based on fixed relations per \eqref{pow}. Notably, the weight matrix $\bbW^{\psi}$ for the final layer can directly adapt to a varying topology by directly updating the $\{\epsilon_{ij}\}$ values. This way, topology-adaptivity is conveniently achieved at the testing phase, and the training dataset does not need to include samples from varying topology. Hence, our layer design ensures high consistency across all PF variables while having the full topology flexibility to model unseen topology scenarios even with the simple fixed-topology training.
\end{remark}

To obtain the trainable GenNN parameters $\{\bbW^1, \bbW^2, \bbb \}$, the Euclidean distance can be employed to construct the loss function.
Based on the input samples, we can calculate the errors between the actual $\bbf(\bbx)= [\bbrho;~\bbpi]$ and $\bbz^2$ generated by the GenNN. 
In addition, the loss function can include  the errors between the actual $[\bbz^{pf}; \bbz^{inj}]$ and the outputs $[\bbz^3;\bbz^4]$ to ensure consistency among all PF variables, given by
\begin{align}
    \mathcal{L}({\bbW^1, \bbW^2, \bbb})= &\| \bbf(\bbx) - \bbz^{2}  \|^2_2 \nonumber \\
    &+\lambda \left\| [\bbz^{pf}; \bbz^{inj}] -  [\bbz^{3}; \bbz^{4}] \right\|^2_2 \label{update}
\end{align}
To balance the two error terms, we set $\lambda>0$ as a regularization hyperparameter.
The choice of $\lambda$ can affect the prediction of the PF variables, as a larger $\lambda$ leads to higher consistency and accuracy in the last two layers.
The proposed GenNN design along with the weighted loss function can ensure a consistent match of the PF equations, which will be demonstrated by extensive numerical tests.

\begin{figure}[t!]
    \vspace{-1.5em}
	\centering
	\subfloat[]{\includegraphics[scale=0.29]{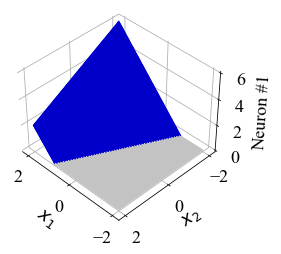}\label{ReLU_a}}
    \subfloat[]{\includegraphics[scale=0.29]{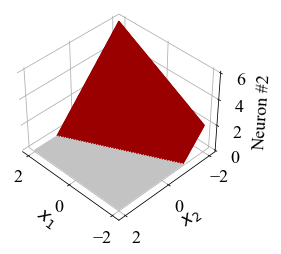}\label{ReLU_b}}
	\subfloat[]{\includegraphics[scale=0.29]{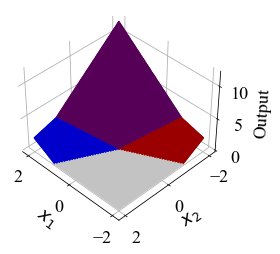}\label{ReLU_c}}
	\caption[]{\small Each ReLU-based hidden neuron (a)-(b) produces a PWL function with two linear regions. The sum of the two neurons (c) becomes another PWL function with four linear regions due to the combination of two neurons' activation status.}\label{ReLU_fig}
\end{figure}

By using the ReLU activation in \eqref{NN_pf1},  the first two layers extend the linearized model in \eqref{linearf} to a PWL function. Note that each hidden neuron $z^1_k$ in \eqref{NN_pf1} essentially uses the activation status to divide the total input space into two different linear regions, forming a PWL relation. To this end, consider a $2\times 1$ input vector $\bbx = [x_1;x_2]$ with two hidden neurons. As illustrated in Figs.~\ref{ReLU_fig}(a)-(b), each of the two hidden neurons produces a PWL relation with two linear regions. The linear combination of these two outputs further divides into multiple linear regions, as shown by Fig.~\ref{ReLU_fig}(c), consisting of four linear regions per the combination of the two ReLU functions' activation status. By using multiple hidden neurons in \eqref{NN_pf1}, our proposed GenNN can express a complex PWL model.

Clearly, the expressiveness of the GenNN depends on the total number of linear regions produced by the ReLU-based hidden layer,  making it important to select the number of neurons. As the number of linear regions scales with the combination of all ReLUs' activation, it can go as high as $2^K$ for a given $K$ neurons. 
Thus, a large enough $K$ has to be used to ensure the resultant PWL model can achieve an accurate approximation. Nonetheless, there is a trade-off between the accuracy and the model complexity by varying $K$, as the total number of GenNN parameters to be trained also increases with $K$. In addition, a high-complexity may lead to a lack of generalizability, namely a loss of accuracy with testing samples that have not been used for training. This is related to the \textit{curse of dimensionality}, a well-known paradox in statistical learning~\cite{trunk1979problem}. Hence, selecting an appropriate $K$ would be critical to ensure a good balance between expressiveness and generalizability. 
We provide insights on this selection here.

\begin{figure}[t!]
	\begin{center}
		\includegraphics[width=0.8\linewidth]{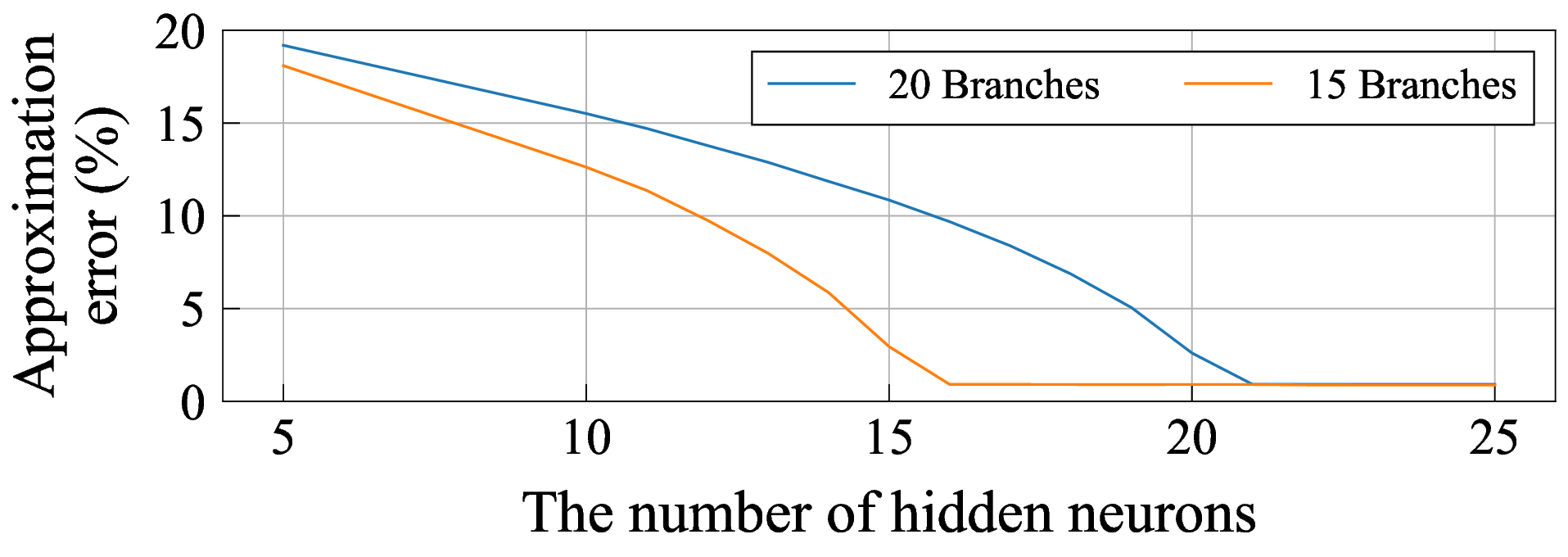}
		\caption{\small Error performance versus the number of hidden neurons for the IEEE 14-bus system with either 20 or 15 branches.} \label{SelectionQ}
	\end{center}
\end{figure}

\begin{remark}{(Selecting the number of hidden neurons.)} \label{rmk:numberK}
As the dimensions of the GenNN's inputs/outputs depend on $N$ and $L$, the numbers of buses/lines, we expect the number of hidden neurons $K$ should scale with them to attain sufficient accuracy.
To choose $K$, we have tested the approximation performance of the first two GenNN layers using the IEEE 14-bus system with either 20 or 15 lines, with the latter having 5 lines deleted from the original test case.
Fig.~\ref{SelectionQ} plots the average approximation errors versus $K$ for both systems, which suggests a good choice for $L=15$ would be $K=14$ and for $L=20$ would be $K=19$. We have performed similar tests in larger systems, and our experimental experiences confirm that $K\approxeq(2N/3+L/3)$ would lead to good accuracy.
\end{remark}

%Needs of clustering perspective of the scalability
Thus, the number of hidden neurons $K$ needed for our GenNN grows with the system size. Assuming the number of lines $L\sim\ccalO(N)$ as in typical power systems \cite{birchfield2016statistical}, this choice of $K$ is also in $\ccalO(N)$. With fully-connected weight matrices $\{\bbW^1, \bbW^2\}$,  the total number of parameters is $\ccalO((N+L)K)$ which becomes $\ccalO(N^2)$.
This quadratic growth rate inspires us to consider sparsifying the weight matrices by partitioning the system into fixed-sized areas.

%large-scale systems require an inevitably greater number of hidden neurons.%ReLU functions.
%The fully connected weight matrices, whose size is proportional to the system inputs and hidden neurons, incur a significant computational burden.
%This practical challenge hinders the scalability of GenNN.
%To preserve it, the ensuing section will tackle the challenge by sparsifying the weights through the system partitioning method.

\subsection{Low-Complexity Sparse GenNN via Area Partitioning}
%First: Why using partitioning is a good approach?
%How to do partitioning?
%decouplde network
%Considering partitioning is useful for sparsification
%Partiiion of the system gives a good route for doing sparsification

To reduce the parameter complexity order, our idea is to consider fixed-size subnetworks by partitioning into multiple areas. The number of parameters for the area-specific GenNN scales with its own size and does not grow with the overall system size. Note that in the PF equations \eqref{powerflow}, the line flows $\{P_{ij},Q_{ij}\}$ depend only on the voltages and angles of the incident buses. In contrast to fully-connected weight matrices $\{\bbW^1, \bbW^2\}$, our area-based approach constrains each submatrix to the dimensions of subnetworks. For example, for an $N$-bus system divided into areas of $N'=25$ buses, the number of weight parameters becomes $\ccalO((N')^2 (N/N'))\sim\ccalO(N\cdot N')$, which is linear in $N$ for fixed $N'$. This low-complexity design is crucial for large-system applications, as the computational and memory demands scale nicely with the overall grid size.

\begin{figure}[t!]
	\centering
	\subfloat[]{\includegraphics[scale=0.35]{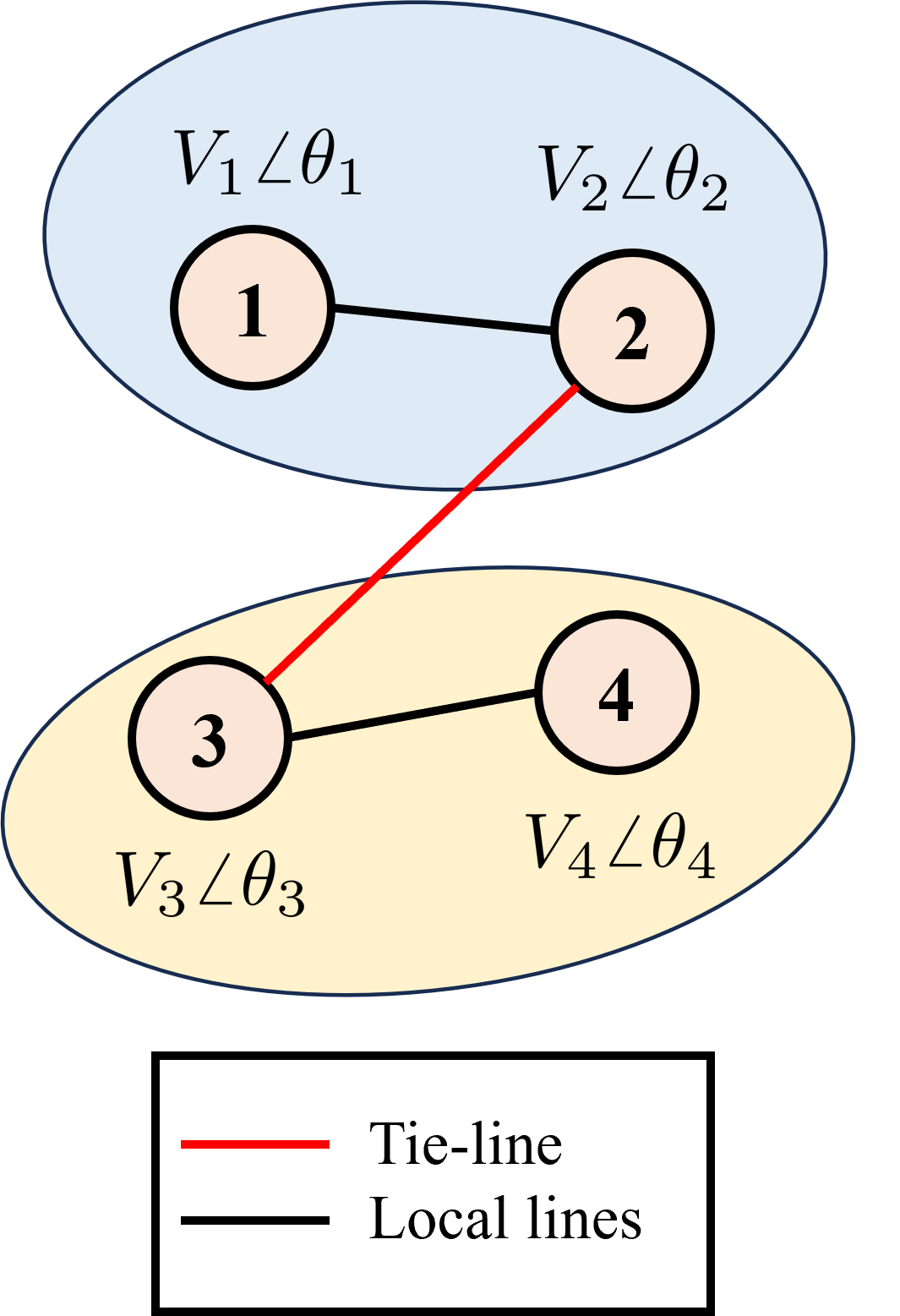}}
	\qquad
	\subfloat[]{\includegraphics[scale=0.35]{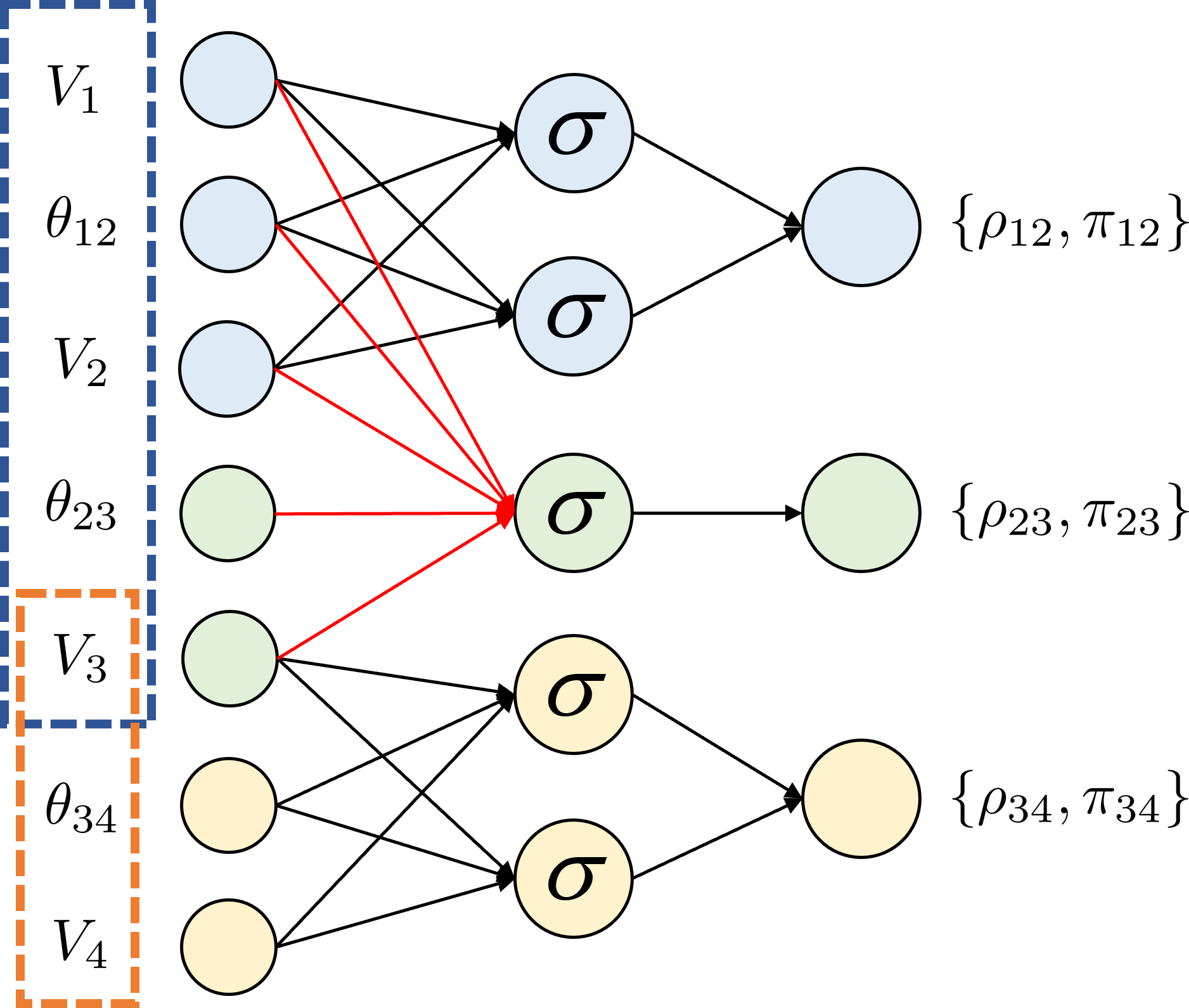}}
	\caption[]{\small (a) A 4-bus system with two areas and a tie-line; and (b) its sparse GenNN. Two individual GenNNs (blue and yellow) are separately formed for the two areas, while the tie-line is incorporated into the top GenNN by introducing more input variables ($\theta_{23}$ and $V_3$).}
    \label{sparseGenNN}
\end{figure}

To illustrate the idea, consider the example system in Fig.~\ref{sparseGenNN}. This 4-bus system is split into two areas with a tie-line connecting them. Within the top area, comprising buses 1 and 2, the line flows $\{P_{12},Q_{12}\}$ depend solely on the local bus states; and likely for the other area. Ignoring the tie-line, two GenNN models could be separately built, one for each area, with the overall weight matrices adopting a block-diagonal structure similar to the PF dependency. Thus, area partitioning not only enforces a fixed parameter order per area size, but also aligns the resultant NN weight with the underlying PF dependency, thereby improving the model performance.

%Second: joint training with tie lines
%Modified version with joint training, joint training at some point needs to mention
Nonetheless, the presence of tie-lines requires connecting the decoupled GenNNs among two neighboring areas.  This tie-line coupling can be conveniently incorporated by expanding the area's inputs with the end bus from the neighboring area. As shown in Fig.~\ref{sparseGenNN}, the tie-line between buses 2 and 3 uses the inputs from both of the two areas. Thus, to predict the line flows $\{P_{23},Q_{23}\}$, the top part of GenNN can be modified by adding the inputs $\{V_3, \theta_{23}\}$.  This modification could create a few additional entries to $\{\bbW^1, \bbW^2\}$, the weight matrices remain predominantly block-diagonal. 
For small systems, these sparse $\{\bbW^1, \bbW^2\}$ can be similarly trained as the case of full matrices earlier, by keeping the zero entries during the training process. When the system size increases, the area-based submatrices can be separately trained using the inputs/outputs within each of the areas including the tie-lines, as discussed later in Remark \ref{rmk:scale}. Either way, the area partitioning-based sparsification still enforces a fixed parameter complexity within each block, thereby significantly reducing the total number of trainable parameters.

%Third: Spectral clustering
\begin{comment}
Since the level of coupling in GenNN depends on the tie-lines, the area partitioning should be designed to reduce the total number, or generally speaking, the \textit{electrical connectivity}, of the tie-lines.  
To this end, we propose to perform the \textit{spectral clustering} on the underlying graph of the power network, which can effectively reduce the inter-area connectivity while producing clusters of high intra-area connectivity \cite{ng2001spectral}. Specifically, we assign a weight parameter for each branch $(i,j)$ using its series impedance, as given by $\lambda_{ij}:=\frac{1}{|z_{ij}|}$ with $z_{ij} = 1/y_{ij}$. This weight choice of $\lambda_{ij}$ follows from the so-termed \textit{electrical distance} between two neighboring buses~\cite{sanchez2014hierarchical}, as a very small line impedance implies that the two end buses are electrically very close. To minimize the total electrical distances, or weights, of the tie-lines, the spectral clustering works by performing the spectral analysis of the weighted graph Laplacian matrix $\bbLambda \in \mathbb R^{N\times N}$, defined by 
\end{comment}

Since coupling in the GenNN depends on tie-lines, the area partitioning should aim to minimize the overall \textit{electrical connectivity} of the tie-lines. To this end, we employ spectral clustering on the underlying power network graph, which reduces inter-area connectivity while preserving strong intra-area connectivity \cite{ng2001spectral}. Specifically, each branch $(i,j)$ is assigned a weight $\lambda_{ij}:=\frac{1}{|z_{ij}|}$ with $z_{ij} = 1/y_{ij}$ reflecting the \textit{electrical distance} between neighboring buses \cite{sanchez2014hierarchical}. To minimize the total electrical distances of the tie-lines, the spectral clustering works by performing the spectral analysis of the weighted graph Laplacian matrix $\bbLambda \in \mathbb R^{N\times N}$, defined by 
\begin{align}\nonumber
    \bbLambda_{ij} &= \begin{cases}
    \textstyle \sum_{j=1}^n \lambda_{ij}, & \text{if $ i= j$,}\\
    -\lambda_{ij}, & \text{if $(i,j)\in \mathcal{L}$,}\\
    0, & \text{otherwise.}
    \end{cases} 
\end{align}
Consider the eigen-decomposition of the symmetric matrix $\bbLambda= \bbR\bbS\bbR^{\mathsf T}$ with the diagonal matrix $\bbS$ having eigenvalues $s_1\leq s_2\leq \cdots \leq s_N$ and $\bbR=[\bbr_1; \bbr_2;\cdots; \bbr_N]$ containing the eigenvectors as the columns.
Since $\bbLambda$ is a graph Laplacian, it is known that the smallest eigenvalue $s_1=0$ with $\bbr_1$ related to the all-one vector \cite{zhu2012sparse}, thereby ``equally'' represents all nodes.
Generally speaking, eigenvectors with smaller eigenvalues capture global connectivity, while those with larger eigenvalues represent local connectivity.
For example, if the second smallest eigenvalue $s_2=0$, the graph has to consist of two disconnected areas. By selecting the $C$ smallest eigenvalues after $s_1$ (i.e., $s_2,\cdots,s_{C+1}$), the corresponding eigenvectors in $[\bbr_2;\cdots; \bbr_{C+1}] \mathbb R^{N\times C}$ serve as features to partition the nodes into up to  $2^C$ areas via {\it{k}}-means clustering.
Although an unweighted Laplacian with $\lambda_{ij}=1$ can be used, our experiments show that the weighted Laplacian based on electrical distance better reduces PF coupling among areas. Upon training the sparse GenNN, we obtain a concise yet low-complexity PWL approximation for the system-wide PF model.

\begin{remark}{(Training for large-scale systems)}\label{rmk:scale}
When the system size is large with e.g. $N\geq 1000$, it could become very computationally demanding to 
allocate all the weight parameters into the computer memory for training. One solution to improve the memory efficiency is to separately build the GenNN for each area, using the inputs/outputs within the area. Note that every tie-line could be assigned to one of the connected areas, which completely eliminates the PF coupling. For the network in Fig.~\ref{sparseGenNN}, we could separately train the two parts of GenNN, as marked by the dashed boxes.  
While this simple decoupled training could lose some accuracy compared to a joint training among all parameters, our numerical experiments have suggested the former still attains a very high approximation performance at a significant reduction of memory needs.   
\end{remark}

\section{GenNN for MILP-based Grid Optimization} \label{sec:MILP_opt}
We formulate our GenNN-based PWL-PF model into a mixed-integer linear (MIL) form, which will be utilized by a variety of grid optimization problems later on.
Recall that for the GenNN model \eqref{NN_pf}, the only nonlinearity lies in the first layer in \eqref{NN_pf1}.
Thus, it suffices to consider the reformulation of the ReLU functions therein only. To this end, we have 
\begin{align}
    \bbz^1 = \sigma(\bbW^1 \tdbbx + \bbb) =  (\bbW^1 \tdbbx + \bbb) \odot \bbbeta \label{eq:prod}
\end{align}
where $\bbbeta$ is an auxiliary binary vector indicating the ReLU activation status, while $\odot$ denotes the entry-wise vector multiplication.
This representation allows us to adopt the well-known \textit{big-M tightening method} for an equivalent linear reformulation~\cite{griva2009linear}. As detailed soon, it is possible to determine the upper/lower bounds of the continuous vector $ (\bbW^1 \tdbbx + \bbb)$ to express \eqref{eq:prod} using four linear inequalities. Note that the complexity of the resultant MIL formulation can be significantly reduced by: i) selecting very tight upper/lower bounds; and ii) eliminating the subset of inequalities that are inactive. We will discuss how to design these two very important aspects specifically for formulating the NN-based relation in \eqref{eq:prod}.

\begin{comment}
The tightness of the upper/lower bounds on the continuous vector crucially affects the computation complexity of the resultant optimization~\cite{fattahi2018bound}.  For example, these bounds could be determined based on the range of the input vector, namely $\tdbbx\in [\underline{\bbx},\bar{\bbx}]$. The latter is easy to obtain for our grid optimization problems, by using the system operating limits on voltage/angle variables as detailed later. Using the range $[\underline{\bbx},\bar{\bbx}]$, one could bound $(\bbW^1 \tdbbx + \bbb)$ by casting some optimization problems; see e.g.,~\cite{kody2022modeling}. Nonetheless, solving the resultant optimization problems inevitably incurs additional computation time.
Instead of using this optimization step, recent work~\cite{anderson2020strong} has developed a direct approach for this bounding specifically for the linear form of $ (\bbW^1 \tdbbx + \bbb)$.
\end{comment}

The tightness of the bounds on the continuous vector affects the computational complexity of the resultant optimization~\cite{fattahi2018bound}. For example, these bounds could be determined from the input range $\tdbbx\in [\underline{\bbx},\bar{\bbx}]$, derived from system operating limits. Although one may bound $(\bbW^1 \tdbbx + \bbb)$ via additional optimization (e.g., \cite{kody2022modeling}), this incurs extra computation. Instead, recent work~\cite{anderson2020strong} has developed a direct bounding approach for the linear form of $(\bbW^1 \tdbbx + \bbb)$.
Let us form two matrices $\{\bbL, \bbU\} \in \mathbb R^{N\times K}$ according to the sign of the entries in $\bbW^1$:%, as given by
\begin{align}
    \{L_{nk},U_{nk}\} =\begin{cases} \{\underline {x}_n, \bar {x}_n \} & \text{if $ W^1_{kn} \geq 0$}\\
    \{\bar {x}_n, \underline {x}_n\} & \text{if $ W^1_{kn} < 0$} \end{cases}.
\end{align}
Using these two matrices, we can express the $k$-th entry in \eqref{eq:prod}, namely $z_k^1$, via four linear inequalities: 
\begin{subequations}\label{MILP}
\begin{align}
    0 &\leq z^1_k \leq (\bbW^1_{k:} \bbU_{:k}\! +\! b_k) \beta_k,\label{beta0} \\
    \bbW^1_{k:}\tdbbx\!+\!b_k &\leq z^1_k \leq \bbW^1_{k:} (\tdbbx\! -\! \bbL_{:k}(1\!-\!\beta_k))\!+\! b_k \beta_k,\label{beta1}
\end{align}
\end{subequations}
where $ \bbW^1_{k:}$ denotes the $k$-th row of matrix $ \bbW^1$.
Basically, if the term $\bbW^1_{k:} \tdbbx+b_k > 0$, then the inequalities $z^1_k\geq\bbW^1_{k:}\tdbbx\!+\!b_k$ in \eqref{beta1} and $z^1_k\leq (\bbW^1_{k:} \bbU_{:k}\! +\! b_k) \beta_k$ in \eqref{beta0} would force $\beta_k=1$, and thus $z^1_k=\bbW^1_{k:} \tdbbx+b_k$ holds under \eqref{beta1}.
Otherwise, if $\bbW^1_{k:} \tdbbx+b_k < 0$, then the inequalities $z^1_k\geq0$ in \eqref{beta0} and $z^1_k \leq \bbW^1_{k:} (\tdbbx\! -\! \bbL_{:k}(1\!-\!\beta_k))\!+\! b_k \beta_k$ in \eqref{beta1} would force $\beta_k=0$, and thus $z^1_k=0$ holds under \eqref{beta0}.
It is shown in \cite{anderson2020strong} that these four linear inequalities based on the range of $\tdbbx$ can attain a much tighter big-M reformulation over traditional bounding approaches. This can greatly improve the solution efficiency of the resultant MILP; see e.g., \cite{anderson2020strong} for more discussions.

\begin{comment}
In addition to facilitating the bounding, the range of $\tdbbx$  can be used for directly determining certain $\beta_k$'s, and accordingly the activation status of some inequalities in \eqref{MILP}. Those inactive inequalities can be eliminated in the resulting MILP, further improving the computation efficiency. 
Note that if the ReLU activation status for the $k$-th neuron $z_k^1$ remains unchanged for the entire range of $\tdbbx$, then we can directly set the value of $\beta_k$ to be either 0 or 1.
Specifically, if $\bbW^1_{k:} \bbU_{:k} + b_k < 0$, then $\beta_k=0$ and thus $z_k^1=0$ have to hold for any $\tdbbx$ in order to satisfy \eqref{beta0}. This way, the inequalities in \eqref{beta1} are no longer active. 
Similarly, if $\bbW^1_{k:} \bbL_{:k} + b_k \geq 0$, then $\beta_k=1$ and thus $z_k^1 = \bbW^1_{k:}\tdbbx+b_k$ have to hold for any $\tdbbx$ to satisfy \eqref{beta1}, making the inequalities in \eqref{beta0} inactive.
\end{comment}

In addition to facilitating the bounding, the range of $\tdbbx$ can directly determine certain $\beta_k$'s and eliminate inactive inequalities, thereby improving computational efficiency. Note that if the ReLU activation status for $z_k^1$ remains unchanged over the entire range of $\tdbbx$, we can directly set $\beta_k$ to either 0 or 1. Specifically, if $\bbW^1_{k:} \bbU_{:k} + b_k < 0$, then $\beta_k=0$ and $z_k^1=0$ for all $\tdbbx$, rendering the inequalities in \eqref{beta1} inactive. Similarly, if $\bbW^1_{k:} \bbL_{:k} + b_k \geq 0$, then $\beta_k=1$ and $z_k^1 = \bbW^1_{k:}\tdbbx+b_k$ for all $\tdbbx$, making the inequalities in \eqref{beta0} inactive.
Hence, we define two subsets of the neurons $k\in\mathcal K:=\{1,\dots,K\}$: 
\begin{subequations}\label{ReLUPrune}
\begin{align}
    \mathcal {K}_0 &= \set[\Big] {k \in \mathcal K \given \bbW^1_{k:}\bbU_{:k} + b_k < 0},\\
    \mathcal {K}_1 &= \set[\Big] {k \in \mathcal K \given \bbW^1_{k:}\bbL_{:k} + b_k \geq 0}.
\end{align}
\end{subequations}
For any neuron $k\in \mathcal K_0$, it suffices to fix $\beta_k=0$ and apply only \eqref{beta0}; and similarly for any $k\in  \mathcal K_1$, it suffices to fix $\beta_k=1$ and apply only \eqref{beta1}.
This way, we can reduce the total number of linear inequalities while maintaining the equivalence of the formulation. Based on our empirical experiences, this step could reduce the number of binary variables for our GenNN representation by around 5\%.

These two steps allow us to reformulate the GenNN-based PF model in \eqref{NN_pf1}-\eqref{NN_pf3} into:
\begin{subequations}\label{PWLPF}
\begin{align}
    \bbW^1_{k:}&\tdbbx\!+\!b_k \leq z^1_k \leq \bbW^1_{k:} (\tdbbx\! -\! \bbL_{:k}(1\!-\!\beta_k))\!+\! b_k \beta_k,\nonumber\\
    &\qquad \qquad \qquad \qquad\qquad~~~~\forall k \in \mathcal K - \mathcal K_0\label{PWLPF1}\\
    0 &\leq z^1_k \leq (\bbW^1_{k:} \bbU_{:k}\! +\! b_k) \beta_k,~\forall k \in \mathcal K - \mathcal K_1\label{PWLPF2}\\
    \beta_k &= \begin{cases} 1,~  k \in \mathcal{K}_1\\
    0,~ k \in \mathcal{K}_0 \end{cases}\label{PWLPF3}\\
    [ \hhatbbrho; \hhatbbpi]&=[\check{\bbrho};\check{\bbpi}]+ \check {\bbJ} \tdbbx + \bbW^2 \bbz^{1}\label{PWLPF4}\\
    \hhatbbgamma &= 2\bbV - \chkbbV\label{PWLPF5}\\
    P_{ij}\!&=\!\hat \gamma_i (\frac{ g_{ij}}{ a^2_{ij}}\!+\!g^{\mathrm {sh}}_i)\!-\!\frac{g_{ij}}{ a_{ij}}\hat \rho_{ij}\!-\!\frac{b_{ij}}{ a_{ij}} \hat \pi_{ij},\!~\forall (i,j) \in \mathcal{L}, \label{PWLPF6}\\
    Q_{ij}\!&\!=\!-\hat \gamma_i (\frac{ b_{ij}}{ a^2_{ij}}\!+\!b^{\mathrm {sh}}_i)\!+\!\frac{b_{ij}}{ a_{ij}}\hat \rho_{ij}\!-\!\frac{g_{ij}}{ a_{ij}} \hat \pi_{ij},\!~\forall (i,j) \in \mathcal{L}. \label{PWLPF7}
\end{align}
\end{subequations}
Basically, we express the first layer \eqref{NN_pf1} into the linear inequalities in \eqref{PWLPF1}-\eqref{PWLPF3}, while the following two linear layers remain unchanged.
Recall that the squared voltage $\bbgamma$ in \eqref{PWLPF5} adopts a first-order approximation. 
This MIL representation can be incorporated into various grid optimization problems to attain an MILP reformulation. We first consider the clairvoyant AC-OPF here, which will be generalized to problems with flexible topology soon. Given the cost function $c_i(\cdot)$ in providing nodal flexibility at bus $i$, the AC-OPF minimizes the total cost while satisfying the operation limits, as given by
\begin{subequations}\label{OPF}
\begin{align}
    \min ~ &\textstyle \sum_{i=1}^N c_{i} (P_{i})\\
    \textrm{s.t.}~ %\quad
        &\underline{V}_i \leq V_{i} \leq \bar{V}_i,~ \underline{\theta}_{ij} \leq \theta_{ij} \leq \bar{\theta}_{ij},~%\label{OPF3}\\
    \beta_k \in \{0,1\}, &\label{OPF1}\\
    &\underline{P}_{i} \leq P_{i} \leq \bar{P}_{i},~\underline{Q}_{i} \leq Q_{i} \leq \bar{Q}_{i},\label{OPF2}\\
    &\underline{P}_{ij} \leq P_{ij} \leq \bar{P}_{ij}, ~ \underline{Q}_{ij} \leq Q_{ij} \leq \bar{Q}_{ij},\label{OPF3}\\
 %   &P^2_{ij}+Q^2_{ij} \leq (\bar{S}_{ij})^2,\label{OPF4}\\
    &\eqref{powerinj}~\textrm{and}~\eqref{PWLPF}.&\label{OPF5}
\end{align}
\end{subequations}
Note that this generic AC-OPF formulation uses $P_i$ and $Q_i$ to represent bus $i$'s total injection flexibility from both generation and demand.
The constraints in \eqref{OPF1} represent the safety limits for the voltage/angle variables and set up the binary activation variable.  Moreover, the power limits in \eqref{OPF2} can account for the presence of non-controllable resources such as the fixed demand, in addition to the resource limits.
Furthermore, the line flow constraints in \eqref{OPF3} are used to mitigate line overloading. While for simplicity only active/reactive flows are considered here,  it is also common to include a constraint on apparent power flow limit. This could be possibly done by either extending our GenNN model to predict apparent line flows or using a quadratic inequality like $P^2_{ij}+Q^2_{ij} \leq (\bar{S}_{ij})^2$. 
Finally, the inclusion of \eqref{powerinj} ensures the nodal power balance with given line connectivity in $\epsilon_{ij}$'s. 
Clearly, the problem \eqref{OPF} is an MILP thanks to the linear reformulation for the PWL-PF as in \eqref{PWLPF}. 
By allowing the line status $\{\epsilon_{ij}\}$ in \eqref{powerinj} to be decision variables, we can extend this AC-OPF formulation to general grid optimization problems with flexible topology. We present two such problems to demonstrate the importance of our topology-aware GenNN design for power flow representation.

\subsection{Optimal Transmission Switching}

The AC optimal transmission switching (OTS) problem aims to determine the line switching in addition to the nodal flexibility in AC-OPF \cite{fisher2008optimal}. Hence, the OTS problem is capable of increasing the grid operational flexibility. Nonetheless, solving OTS, especially the one based on AC power flow, is challenged by the presence of integer decision variables used for selecting the line status. Our proposed PWL-PF model can facilitate the reformulation of AC-OTS into MILP. To this end,    
let us denote $\hhatP_{ij} := \epsilon_{ij}  P_{ij}$ and $\hat{Q}_{ij} := \epsilon_{ij}  Q_{ij}$ in \eqref{powerinj}, where the line status $\epsilon_{ij}$ becomes a binary decision variable in OTS. To deal with these multiplication terms, we can similarly apply the big-M tightening method for an MIL representation by adding additional constraints to the AC-OPF \eqref{OPF}, as below: 
\begin{subequations}
\begin{align}
    \min \quad &\textstyle \sum_{i=1}^N c_{i} (P_{i})\\
    \textrm{s.t.} \quad
    &\eqref{OPF1}-\eqref{OPF5}, \epsilon_{ij} \in \{0,1\}\\
    &{\underline{ P}}_{ij}  \epsilon_{ij} \leq \hat{ P}_{ij} \leq {\bar{ P}}_{ij}  \epsilon_{ij}, \label{sw0_P}\\
    &{\underline{ Q}}_{ij}  \epsilon_{ij} \leq \hat{ Q}_{ij} \leq {\bar{ Q}}_{ij}  \epsilon_{ij}, \label{sw0_Q}\\
    P_{ij} + &{\bar{ P}}_{ij} ( \epsilon_{ij}-1) \leq \hat{ P}_{ij} \leq  P_{ij} + {\underline{ P}}_{ij} ( \epsilon_{ij}-1), \label{sw1_P}\\
    Q_{ij} + &{\bar{ Q}}_{ij} ( \epsilon_{ij}-1) \leq \hat{ Q}_{ij} \leq  Q_{ij} + {\underline{ Q}}_{ij} ( \epsilon_{ij}-1). \label{sw1_Q}\\
    &\textstyle \sum_{(i,j) \in \mathcal{L} } ~\epsilon_{ij} \geq L-\alpha\label{alpha}
\end{align}
\end{subequations}
where the last constraint \eqref{alpha} sets a total switching budget of no more than $\alpha$ lines.
The constraints \eqref{sw0_P}-\eqref{sw1_Q} are the linear formulation for $\hat{ P}_{ij}$ and $\hat{ Q}_{ij}$, by using their respective bounds per the big-M method.
By reformulating the PWL-PF and line switching relations, we cast the AC-OTS  as an MILP, for which efficient solvers will be used.% as shown by the numerical tests soon.

\subsection{Restoration Ordering Problem}
We consider the power system restoration ordering problem (ROP)~\cite{van2011vehicle}, which extends the static OPF problem to a multi-period setting.
The ROP aims to optimize the restoration order of the damaged transmission lines to maximize the energy served to customers while maintaining system stability.
Similar to the OTS problem, the ROP also includes the status of the damaged lines as decision variables, along with the power flow.
However, as the instantaneous restoration budget is usually very small due to limited resources, the restoration process has to span over multiple periods and thus the ROP is considered over a time horizon $ \mathcal{T} := \{1,\dots, T\}$.
Given a restoration budget of $\eta$ lines, the total number of connected lines at  time $t$ is updated  by
\begin{align}
    R_{t} = R_{t-1}+\eta,~\forall t \in \mathcal{T},
\end{align}
with $R_0$ initialized by the post-damage power system. This equality establishes the temporal coupling in the ROP.  

The  ROP aims to determine the binary line status variables $\epsilon_{ij,t}\in\{0,1\}$  per line $(i,j)$, and the percentage of load shedding $x_{i,t}\in [0,1]$ per bus $i$, at every time  $t\in\ccalT$.   
With the objective of maximizing the total energy provided to all customers throughout the restoration process, the ROP can be formulated using the PWL-PF model in \eqref{PWLPF}, as 
\begin{subequations}
\begin{align}
    \max \quad & \textstyle\sum_{t=1}^T \textstyle\sum_{i=1}^N x_{i,t} P_{i,t}\\
    \textrm{s.t.} \quad
    &\eqref{OPF1}-\eqref{OPF5},~\epsilon_{ij,t} \in \{0,1\},~x_{i,t}\in [0,1]\\
    &{\underline{ P}}_{ij.t}  \epsilon_{ij,t} \leq \hat{ P}_{ij,t} \leq {\bar{ P}}_{ij,t}  \epsilon_{ij,t}, \label{ROP1}\\
    &{\underline{ Q}}_{ij,t}  \epsilon_{ij,t} \leq \hat{ Q}_{ij,t} \leq {\bar{ Q}}_{ij,t}  \epsilon_{ij,t}, \label{ROP2}\\
    P_{ij,t} + &{\bar{ P}}_{ij,t} ( \epsilon_{ij,t}-1) \leq \hat{ P}_{ij,t} \leq  P_{ij,t} + {\underline{ P}}_{ij,t} ( \epsilon_{ij,t}-1), \label{ROP3}\\
    Q_{ij,t} + &{\bar{ Q}}_{ij,t} ( \epsilon_{ij,t}-1) \leq \hat{ Q}_{ij,t} \leq  Q_{ij,t} + {\underline{ Q}}_{ij,t} ( \epsilon_{ij,t}-1), \label{ROP4}\\
    &\textstyle \sum_{(i,j) \in \mathcal{L} }~\epsilon_{ij,t} \leq R_t, \quad \label{ROP5}\\
    &R_{t} = R_{t-1}+\eta, ~\epsilon_{ij,t-1} \leq \epsilon_{ij,t},%\label{ROP6}\\
    %&
    ~\epsilon_{ij,T}=1.\label{ROP7}
\end{align}
\end{subequations}
Note that the initial $\epsilon_{ij,0}$ is given by the post-damage line status, while \eqref{ROP7} sets the final $\epsilon_{ij,T}=1$ to ensure a full restoration. Similar to the OTS, the constraints \eqref{ROP1}-\eqref{ROP4} are used to form the actual line flows $\hhatP_{ij,t}$ and $\hhatQ_{ij,t}$ according the line status $\epsilon_{ij,t}$ per time $t$. By bounding the total number of connected lines in \eqref{ROP5}, the temporal coupling in \eqref{ROP7} ensures that the restoration budget $\eta$ is met and the restored lines will stay connected throughout the restoration.    
Thanks to our proposed GenNN-based PWL-PF representation, the ROP can be again cast as an MILP and will be efficiently solved. % as shown by the ensuing numerical tests. 

\section{Numerical Studies} \label{sec:sim}

We have validated the proposed GenNN modeling approach on the IEEE 118-bus test case~\cite{IEEE_case_ref} and the 6716-bus synthetic grid case~\cite{mateo2024building} for power flow modeling and grid topology optimization tasks. The 6716-bus test case is a synthetic power system developed to represent the geographical coverage of Texas. 
The GenNN training has been performed in PyTorch with Adam optimizer on a regular laptop with Intel\textsuperscript{\textregistered} CPU @ 2.70 GHz, 32 GB RAM, and NVIDIA\textsuperscript{\textregistered} RTX 3070 Ti GPU @ 8GB VRAM. We have formulated the OTS and ROP problems through Pyomo~\cite{hart2011pyomo} and used the Gurobi optimization solver~\cite{gurobi} to solve the resultant MILPs.

To train the GenNN, we generate 10,000 samples based on the actual PF model with the nominal topology by setting $\epsilon_{ij}=1$ for all lines. As emphasized in Remark \ref{rmk:layer}, this fixed-topology based training can still allow for full topology flexibity at the testing/optimization phase, thanks to our layer design, by updating the $\epsilon_{ij}$ values in the final GenNN layer  \eqref{NN_pf4}. The target outputs for training include nonlinear terms $\{\bbgamma, \bbrho, \bbpi\}$ along with the line flows and nodal injections.
For each sample, we generate uniformly distributed voltage magnitudes within the range of [0.94,~1.06] p.u. for the 118-bus case, and of [0.90,~1.10] p.u. for the 6716-bus case, per the voltage limits in each case. Similarly, the voltage angles are randomly varied within $[-\pi/6, \pi/6]$ radians around the initial operating point.
For the reference bus, %namely, {Bus 69} in the 118-bus system or {Bus 111333} in the 6716-bus system, 
we fix its voltage magnitude and angle at the default values.
%Per Remark \ref{rmk:layer}, the training dataset uses solely the original grid topology with $\epsilon_{ij}=1$ for all lines and thus is scalable without the need of varying topology. The flexible topology is conveniently achieved  during the testing or optimization phase using variable $\epsilon_{ij}$'s inside the last GenNN layer in \eqref{NN_pf4}.
The NN parameters have been trained via backpropagation using the loss function \eqref{update} with $\lambda=10$, with a total of $20$e$3$ epochs and a learning rate of $2.5$e$-3$. Note that the $\lambda$ value has been chosen through hyperparameter tuning. We have used 90\% of the dataset for training and the rest 10\%  for testing. The NN modeling results and error performance comparisons presented later on are based on the testing data only.

\subsection{AC Power Flow Modeling}
We consider the AC-PF modeling performance to demonstrate the advantages of our GenNN model over the existing design~\cite{kody2022modeling,chevalier2022global}, which directly predicts all PF variables and is indicated by DirectNN. The DirectNN model is trained using a two-layer NN to predict $\bbz^{pf}$ and $\bbz^{inj}$ from the input voltage variables $\bbV$ and $\bbtheta$.  
We also compare the scalability of the two configurations of our proposed GenNN, the one with fully connected weights (GenNN\_Full), as detailed in our earlier work \cite{cho2023topology}, and the other with the low-complexity GenNN with sparsified weights (GenNN\_Sparse).

\begin{figure}[t!]
	\centering
	\subfloat[Average error]{\includegraphics[scale=0.25]{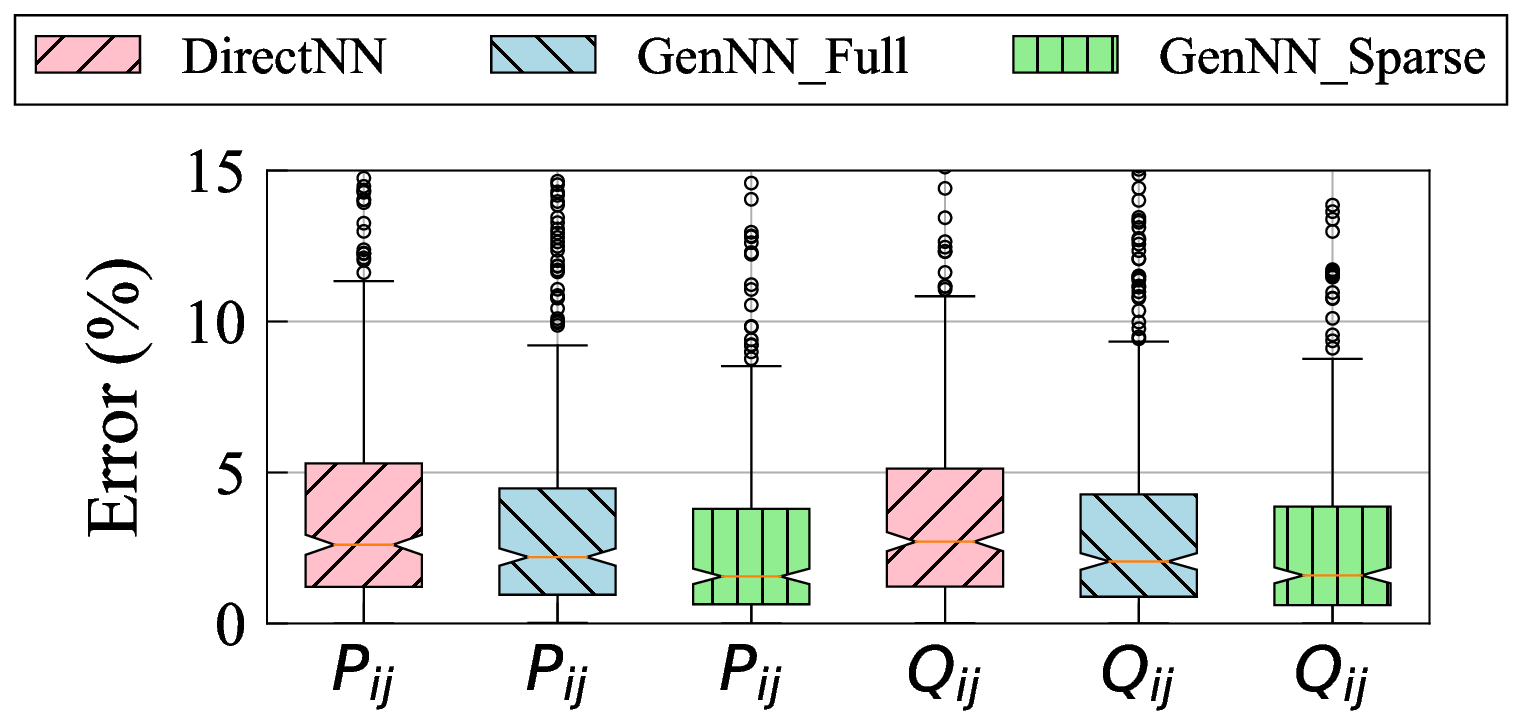}\label{avg_118}}
	\quad
	\subfloat[Maximum error]{\includegraphics[scale=0.25]{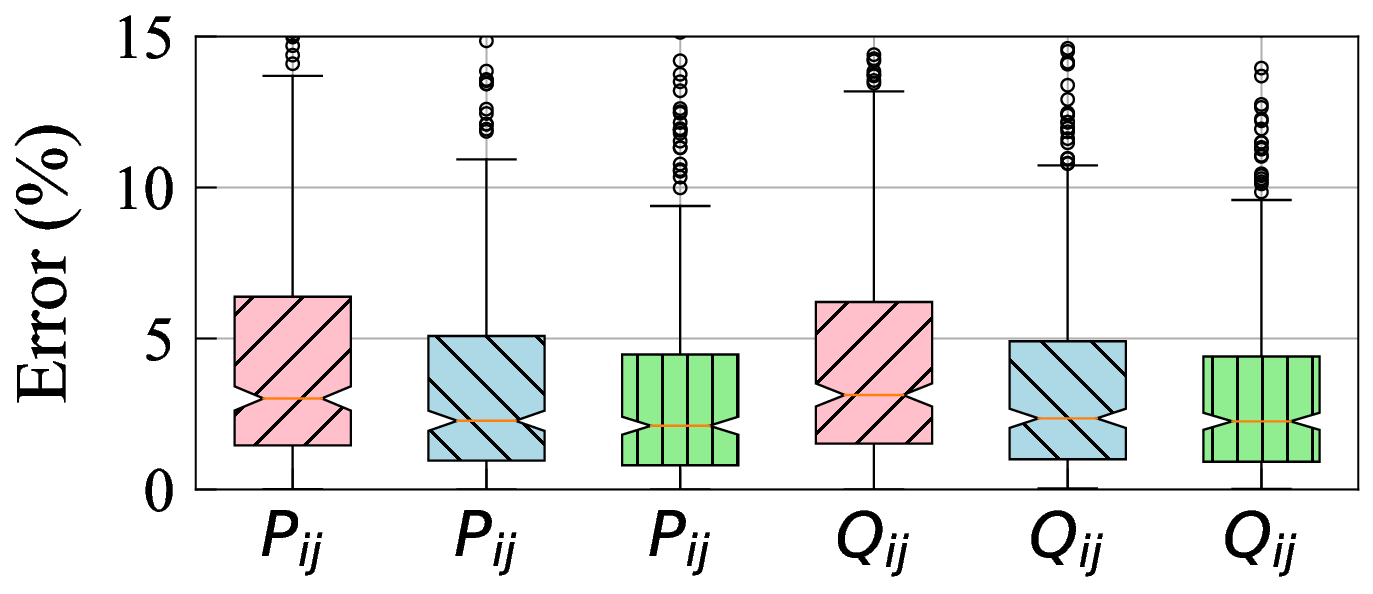}\label{max_118}}
    \quad
    \subfloat[Injection error]{\includegraphics[scale=0.25]{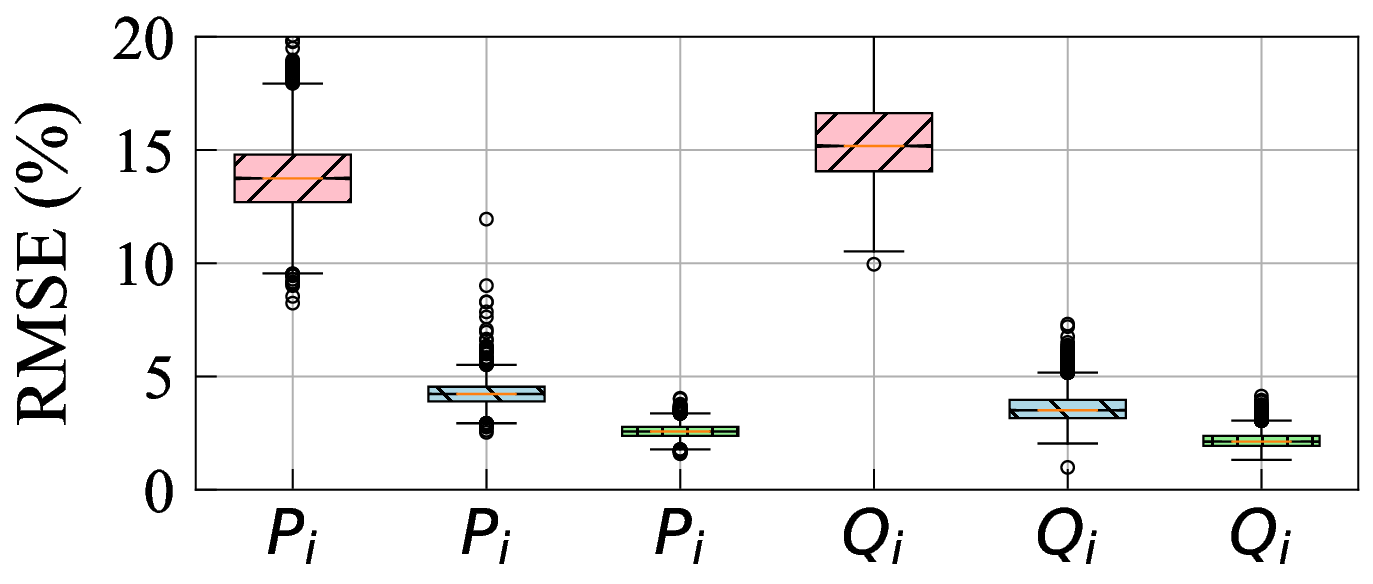}\label{inj_118}}
	\caption[]{\small Comparisons of the (a) average and (b) maximum error in approximating the line power flows and (c) RMSE in predicting the injected power vectors for Direct, GenNN\_Full and GenNN\_Sparse methods using the 118-bus system.}\label{ACPF_118}
\end{figure}

First, we present the approximation error performance using the small 118-bus case. All three NN-based models here use a total of $K=100$ hidden neurons for the trainable layers, in accordance with Remark \ref{rmk:numberK}. To obtain the GenNN\_Sparse model, we partition the system into 5 areas using the $k$-means clustering based upon the top 3 eigenvectors of the Laplacian, and the 100 hidden neurons have been uniformly spread across the 5 areas.
Fig.~\ref{ACPF_118} shows the error box plots for predicting the line flows and the power injections. Fig.~\ref{ACPF_118}(a) and Fig.~\ref{ACPF_118}(b) respectively plot the average and maximum statistics, for the error percentages of active and reactive line flows as normalized by the line capacity.
In addition, Fig.~\ref{ACPF_118}(c) shows the root mean square error (RMSE) in predicting the injected power vectors.
Each box plot indicates the median values as midlines along with the first (Q1) and third (Q3) quartiles as boxes. The horizon bars denote the farthest data points within 1.5 times the interquartile range (Q1-Q3) from the box.

We can observe that compared to DirectNN, the two GenNN models have noticeably reduced the error in approximating both line flows and power injections.
This performance improvement of GenNN is especially significant in terms of approximating the injections $\bbz^{inj}$ as shown in Fig.~\ref{ACPF_118}(c). Since the DirectNN model overlooks the underlying relations between $\bbz^{pf}$ and $\bbz^{inj}$, it is prone to produce power flow patterns that are inconsistent with the nodal power balance in \eqref{powerinj}. In contrast, our GenNN models have been designed to maintain this power flow balance condition thanks to its generative structure between the last two layers in Fig.~\ref{structure}. In addition, Figs.~\ref{ACPF_118}(a)-(b) show our GenNN models have also led to some improvements in approximating the line flows  $\bbz^{pf}$. This is thanks to the GenNN design by embedding the underlying relations from the nonlinear terms $\{\bbgamma, \bbrho, \bbpi\}$ to $\bbz^{pf}$.  Clearly, the proposed generative structure has shown to be effective for the GenNN-based PWL models to maintain the physical relations among PF variables.
Furthermore, the comparisons between the two GenNN models have corroborated the importance of reducing the number of NN parameters, as the  GenNN\_Sparse exhibits a higher accuracy than GenNN\_Full. The advantages of having the  GenNN\_Sparse design will become even more evident in the large system tests as detailed soon.

\begin{figure}[t!]
	\centering
	\subfloat[Average error]{\includegraphics[scale=0.25]{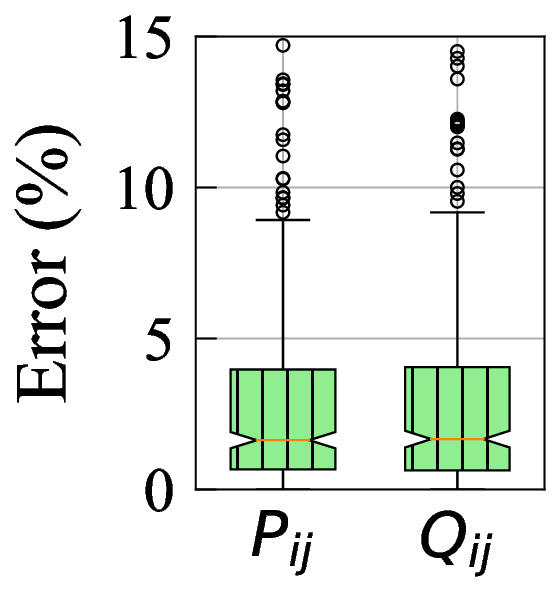}\label{avg_TX}}
	\quad
	\subfloat[Maximum error]{\includegraphics[scale=0.25]{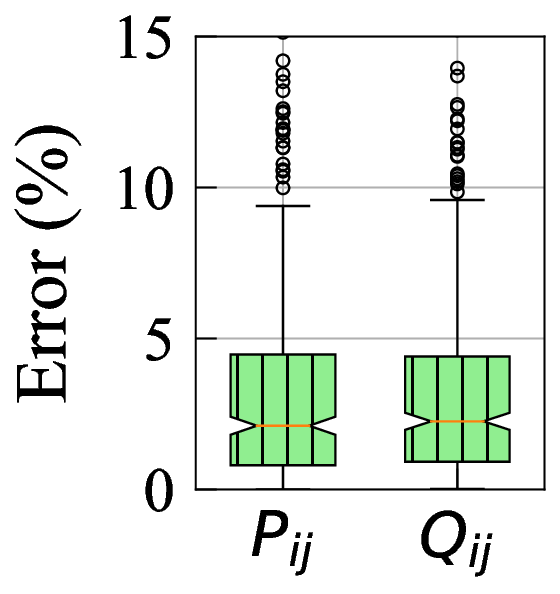}\label{max_TX}}
    \quad
    \subfloat[Injection error]{\includegraphics[scale=0.25]{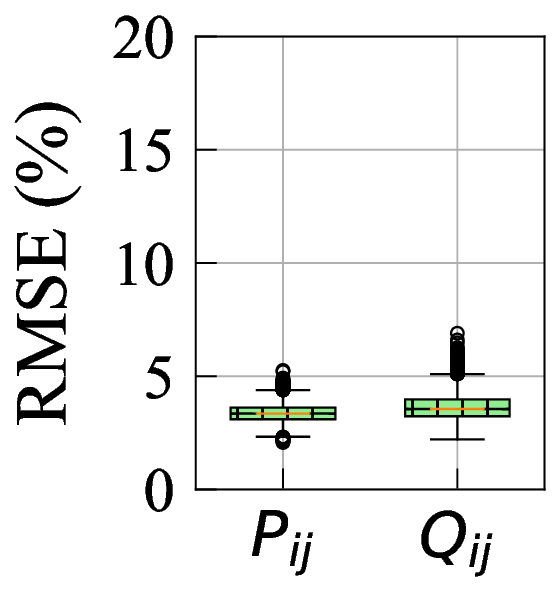}\label{inj_TX}}
	\caption[]{\small Comparisons of the (a) average and (b) maximum error in approximating the line power flows and (c) RMSE in predicting the injected power vectors for GenNN\_Sparse using the 6716-bus system.}\label{ACPF_TX}
\end{figure}

%Only for Sparsified NN
To demonstrate the scalability enhancements of the proposed  GenNN\_Sparse model, we further test it using the large 6716-bus case.
Due to the significant increase in system size, the DirectNN model and even the GenNN\_Full model incur extremely high computational burdens for this large case. The NN training for either of the two models could not be completed within a reasonable time on our computing platform. Therefore, we end up with implementing the GenNN\_Sparse model only here, which can gracefully handle practical systems of a large size as discussed in Remark \ref{rmk:scale}.  
Specifically, we have partitioned this system into a total of 250 areas by using the top 8 eigenvectors of the Laplacian, leading to approximately 27 nodes in each area on average. This is very close to the area size in the 118-bus case partitioning. A total of $K=7500$ hidden neurons has been used for the trainable layers, uniformly spreading across 250 areas. Similar to Fig.~\ref{ACPF_118}, we plot the three error statistics in Fig.~\ref{ACPF_TX}, but only for the GenNN\_Sparse model. 
Although we cannot compare the performance with the other NN-based models, the level of approximation accuracy by the proposed GenNN\_Sparse is observed to be very similar to that in the 118-bus system. Albeit the increase of system size, the error percentages maintain to be mostly below 5\% in the 6716-bus case. This performance guarantee is thanks to the proposed sparse design based on area partitioning, such that the low complexity for the individual areas stays at the same level. To sum up, the numerical results have well supported the benefits of the proposed GenNN model in maintaining PF relations and improving accuracy, as well as the scalability of the sparse GenNN design in reducing the model complexity and computation burdens.
Due to the scalability and high accuracy attained by GenNN\_Sparse, the rest of the simulations will incorporate this model into the topology-related grid optimization tasks.

\subsection{Applications to the OTS}
We implement the proposed GenNN-based PWL-PF model for solving the AC-OTS problem. We will compare it with the DC- and AC-OTS algorithms, both provided by the open-source platform PowerModels.jl ~\cite{8442948}.
%{\cred
Note that due to the nonconvexity of AC-OTS, the solver therein has adopted the combination of Juniper~\cite{juniper} and interior point optimizer (IPOPT), which may suffer from some sub-optimality issues. Nonetheless, its AC-OTS algorithm could be considered as an excellent baseline for our PWL-based method.
%}
To compare across the three OTS methods, we take the line-switching $\{\epsilon_{ij}\}$ solution from each OTS algorithm's outputs, and fix these topology-related decision variables to re-run the AC-OPF problem using the MATPOWER~\cite{zimmerman2010matpower} solver.
This way, we can evaluate the performance of the resultant AC-OPF, in terms of the objective costs (for optimality), as well as the percentages of solver failure and constraint violations (for feasibility).
We use the objective cost of the AC-OPF problem (with no line-switching flexibility) as the basis to normalize the minimum costs of the OTS methods, and thus the latter will attain the cost percentage below 100\%. 
As for the feasibility, the solver failure metrics measure the rates of the follow-up AC-OPF solver returning an infeasible case, while the constraint violation one depends on the occurrences of operational constraint over-limits for the feasible cases only. Both metrics can evaluate the quality of the OTS topology solutions in terms of satisfying the operational limits. In addition to these performance metrics, we also compare the computation time for running each of the OTS algorithms excluding the follow-up AC-OPF solution time.

\begin{figure}[t!]
	\begin{center}
		\includegraphics[scale=0.33]{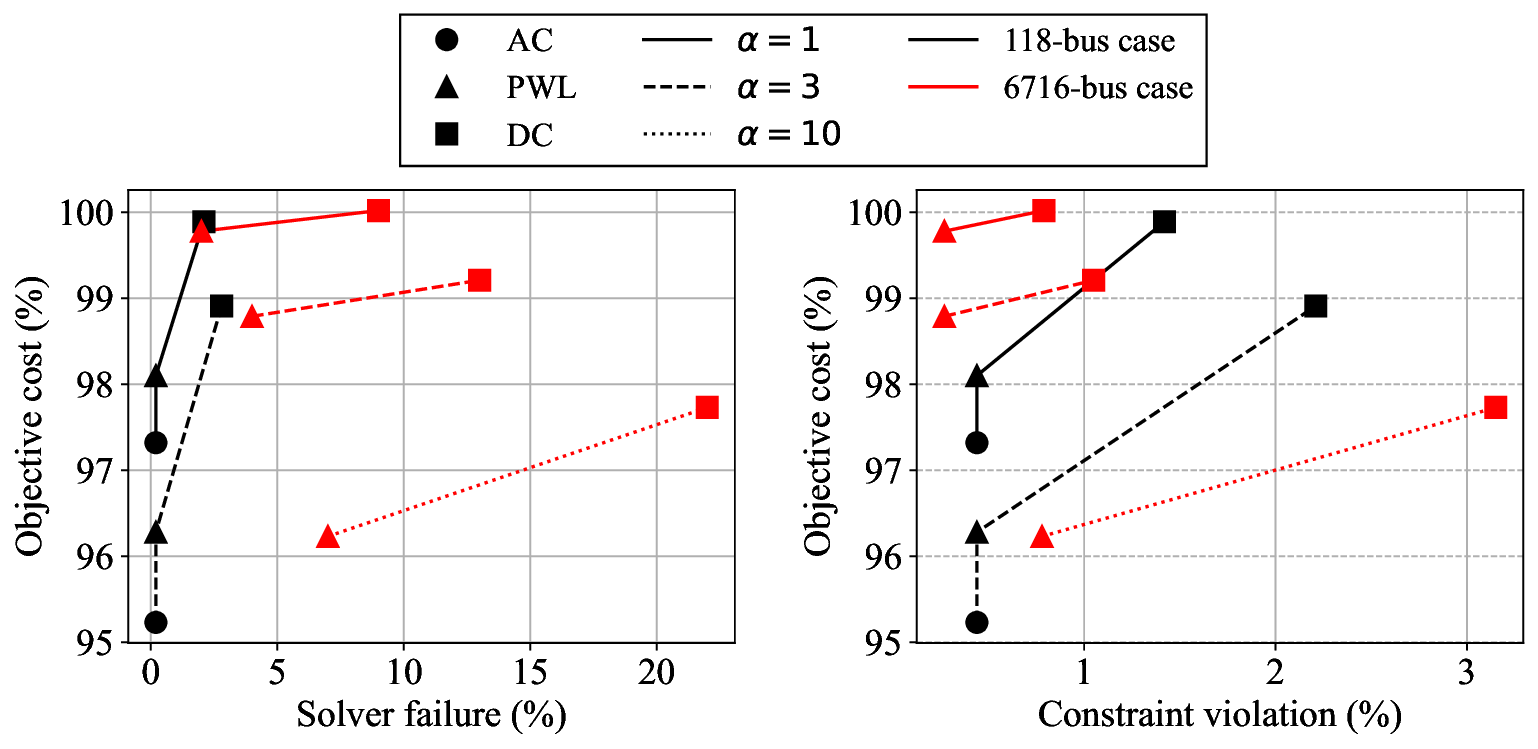}
		\caption{\small Comparisons of the OTS optimality versus feasibility performance using the 118-bus and 6716-bus systems.} \label{OTS_sim}
	\end{center}
\end{figure}

\begin{figure}[t!]
	\begin{center}
		\includegraphics[scale=0.3]{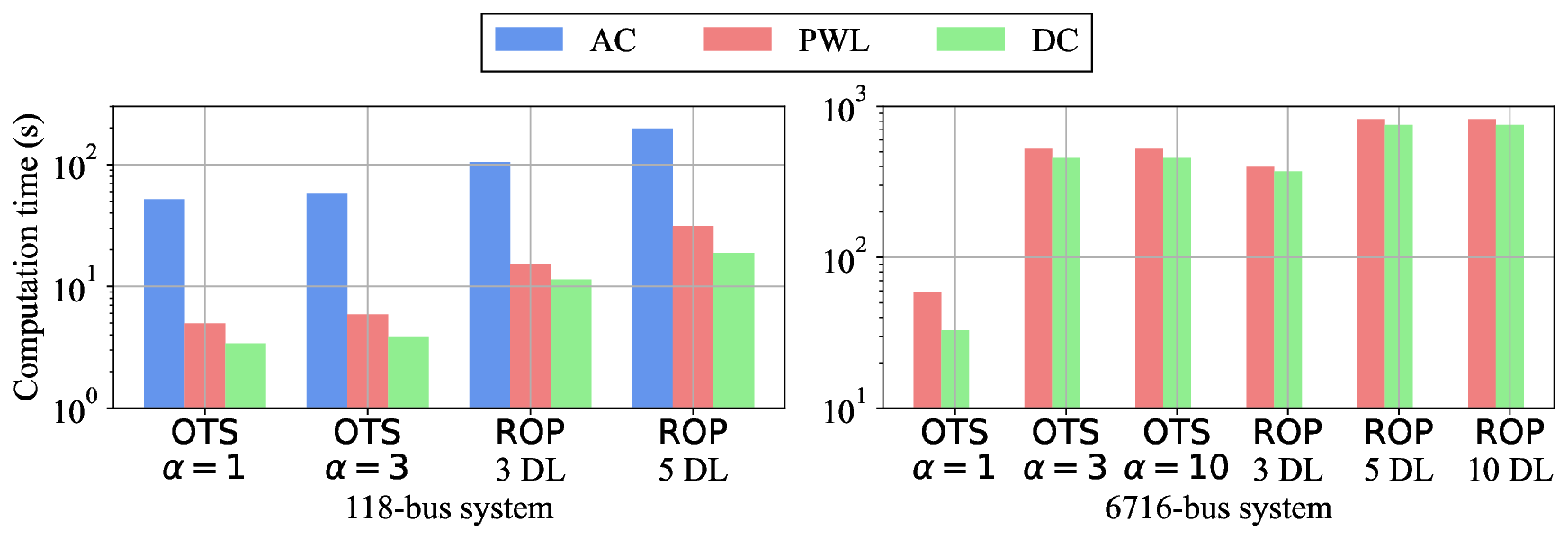}
		\caption{\small Comparisons of the OTS and ROP computation time  using the 118-bus and 6716-bus systems for OTS and ROP.} \label{ComputationTime}
	\end{center}
\end{figure}

We compare the three OTS methods for the 118-bus case in Fig.~\ref{OTS_sim}, which illustrates the trade-off between optimality and feasibility metrics. A switching budget $\alpha$ of either 1 or 3 is considered.
A total of 1,000 PF scenarios by having nodal demands uniformly distributed within [50\%, 200\%] of the nominal values given in~\cite{IEEE_case_ref} have been generated for evaluating  the average performance.
Compared with the AC-OTS, the proposed PWL-OTS approach slightly increases the objective cost by less than 1\% while attaining exactly the same feasibility. The competitive optimality-feasibility performance of the PWL-OTS based topology decisions clearly confirms the high approximation accuracy of our proposed GenNN model. The DC-OTS unfortunately suffers from a poor feasibility performance due to its lack of PF modeling accuracy.
As for computation efficiency, the proposed PWL-OTS achieves the same order of DC-OTS, as shown in Fig.~\ref{ComputationTime}, at approximately one-tenth of the AC-OTS' solution time.

We also perform the OTS comparisons using the 6716-bus system, which includes 135 switchable high-capacity lines. For this large system, it becomes impractical to run the AC-OTS algorithm in PowerModels.jl which takes about 40 mins.
As a result, we compare the proposed PWL-OTS with only DC-OTS, as shown in Fig.~\ref{OTS_sim}. A larger switching budget $\alpha$ of 10 is also considered.
Similar to the previous results, 
the proposed PWL-OTS approach attains a much better performance of optimality and feasibility over DC-OTS. The PWL-OTS's improvements on the feasibility metrics are especially remarkable, thanks to the high accuracy of our proposed GenNN model. Again, its computation complexity is at the same level as the DC-OTS one.    
All these comparisons have validated the suitability of our  GenNN-based PWL power flow model for the topology-optimization task like OTS, as it has gained the near-optimal performance while maintaining high solution feasibility and low computation complexity.

\begin{figure}[t!]
	\begin{center}
		\includegraphics[scale=0.33]{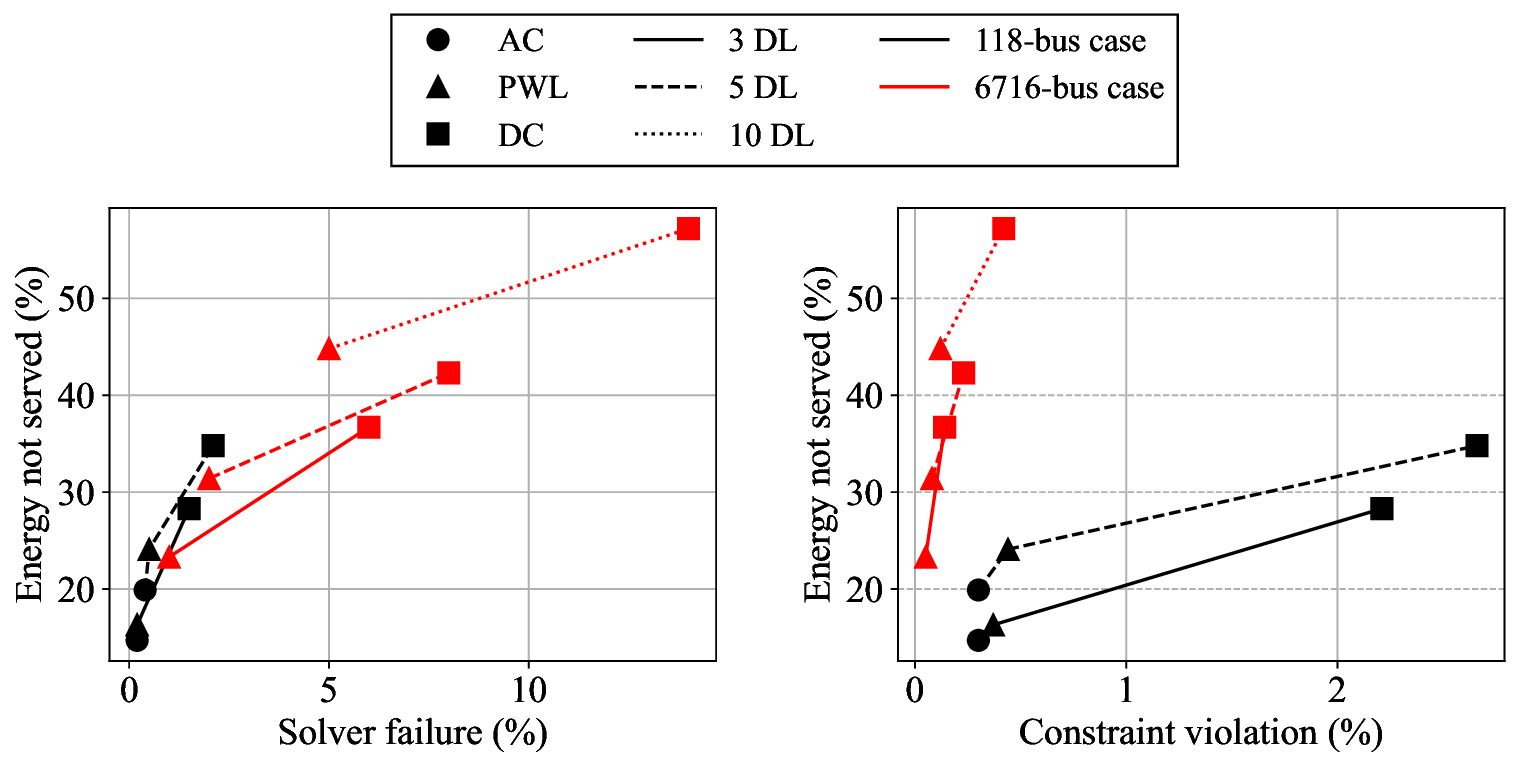}
		\caption{\small ROP comparisons on the 118-bus and 6716-bus systems.} \label{ROP_sim}
	\end{center}
\end{figure}

\subsection{Applications to the ROP}
We utilize the proposed GenNN-based PWL-PF model for solving the AC-ROP problem.
We will compare it with the ROP algorithms using the DC- and AC-PF models, both provided by the open-source platform PowerModelsRestoration.jl~\cite{rhodes2020powermodelsrestoration}.
Note that similar to the AC-OTS, the AC-ROP algorithm is likely to be sub-optimal but still used here as the baseline for comparisons. 
To compare across the three ROP methods, we take the solution of restoration order $\{\epsilon_{ij_t}\}$ for every time $t\in\ccalT$ from each ROP algorithm's outputs, and fix these decision variables to re-run an AC PF-based optimization problem that intends to maximize the total amount of energy served to customers, as given by
\begin{subequations}\label{rop_post}
\begin{align}
    \max \quad & \textstyle\sum_{t=1}^T \textstyle\sum_{i=1}^N x_{i,t} P_{i,t}\\
    \textrm{s.t.} \quad
    & \eqref{OPF1}-\eqref{OPF5}.
\end{align}
\end{subequations}
This way, we can compare the ROP methods based upon the performance in solving the resultant optimization \eqref{rop_post}, in terms of the energy not served (for optimality), as well as solver failure and constraint violations (for feasibility).
For the optimality comparison, we compute the percentage of energy not served to the customers on the damage-affected buses $\mathcal{N}^\times \subset \mathcal{N}$ during the restoration, as given by
\begin{align}
\frac{\textstyle\sum_{t \in \mathcal{T}} \textstyle\sum_{i\in \mathcal{N}^\times} (1-x_{i,t}) P_{i,t}}{\textstyle\sum_{t \in \mathcal{T}} \textstyle\sum_{i\in \mathcal{N}^\times} P_{i,t}}.
\end{align}
The lower this percentage is, the better the optimality performance is.
As for the feasibility, the solver failure and constraint violation metrics are similar to those considered in the OTS.
In addition to these performance metrics, we also compare the computation time for running each of the ROP algorithms.

We first compare the three ROP methods for the 118-bus case in Fig.~\ref{ROP_sim}, which illustrates the trade-off between optimality and feasibility metrics. We set the restoration budget to $\eta=1$ and consider cases with 3 or 5 damaged lines (DLs) which result in $T=3$ or $T=5$, respectively. A total of 1,000 PF scenarios have been similarly generated to compute the average performance.
We observe that compared with the AC-ROP, the proposed PWL-ROP approach slightly suffers in optimality performance, as its percentage of energy not served increases a bit. However, our PWL-ROP maintains the feasibility metrics.
This speaks for the competitive optimality-feasibility performance of the PWL-ROP based topology decisions and confirms the high accuracy of our proposed GenNN model. Meanwhile, the PWL-ROP algorithm achieves a very high computation efficiency, as shown in Fig.~\ref{ComputationTime}. It takes only a sixth of the AC-ROP's solution time. Using the same computation complexity to DC-ROP, our proposed PWL-ROP approaches the performance of AC-ROP.

We also compare it using the 6716-bus system by excluding the AC-ROP due to its very high computational complexity.
Fig.~\ref{ROP_sim} compares the proposed PWL-ROP with DC-ROP. A larger case with 10 DLs is also considered.
These large system-based comparisons again confirm the previous observations and verify the PWL-ROP's excellent optimality-feasibility performance improvements over DC-ROP. But yet, the proposed PWL-ROP's computation time is in the same order as the DC-ROP one.    
Thanks to the low-complexity design of our PWL-PF model, it can greatly simplify the computations for multi-period topology optimization tasks like ROP while attaining an excellent balance between optimality and feasibility.

%Especially as the system size increases, the feasibility of the DC-ROP solution deteriorates significantly due to the %inaccuracy of the DC approximation.

\section{Conclusions} \label{sec:con}
We proposed a novel GenNN-based model that accurately approximates nonlinear AC-PF equations and ensures high consistency among PF variables. Our low-complexity design uses an area-based sparsification strategy to attain linear parameter scaling for large systems. By embedding binary topology variables in the GenNN layers, it can adapt to varying topologies without requiring the generation of such training samples. This resultant PWL power flow approximation is successfully applied to grid optimization tasks such as optimal transmission switching (OTS) and restoration ordering problems (ROP) via an MILP reformulation. Numerical tests on the IEEE 118-bus and 6716-bus synthetic Texas grids corroborate its superior modeling accuracy, as well as the scalability and computational efficiency for topology optimization. 

Future directions include extending our GenNN PF models to distribution and carbon emission flows, enhancing grid monitoring with limited sensing, and further simplifying MILP reformulations in PWL-based grid optimization.

\begin{comment}
We developed a data-driven approach for piecewise linear (PWL) PF modeling, using a novel  NN layer design with ReLU activation to match the generative structure of AC-PF model.
%
The resulting generative NN (GenNN)  models can effectively preserve the consistency among the predicted power variables and integrate the topology decision variables, enabling a mixed-integer linear program (MILP) reformulation for grid topology optimization tasks.
%
We further designed an area-partitioning based sparsification method to reduce the GenNN model complexity.
%
The sparse GenNN models have excellent scalability for large-scale systems, leading to efficient solutions for AC-PF based optimal transmission switching (OTS) and restoration order problems (ROP).
%
Numerical tests on the IEEE 118-bus and the 6716-bus synthetic Texas grid systems have demonstrated the performance improvements of our proposed models in both approximating the AC-PF and expediting topology optimization solutions.
%

Exciting future directions open up for our GenNN-based linearizable PF models, such as extensions to modeling distribution power flow or carbon emission flow. We can also consider general grid monitoring and operation tasks by accounting for limited sensing and control capabilities. Moreover, implementing the PWL-based grid optimization solutions could benefit from reducing the complexity of the MILP reformulation.
\end{comment}

\ifCLASSOPTIONcaptionsoff
\newpage
\fi
	
\bibliographystyle{IEEEtran}
\bibliography{Ref.bib}

\end{document}